\pdfoutput=1
\documentclass[runningheads]{llncs}
\usepackage{physics}
\usepackage{amssymb}
\usepackage{listings}
\usepackage{graphicx}
\usepackage{capt-of}
\usepackage[all]{xy}
\usepackage{tikz}
\usepackage{multirow}
\usetikzlibrary{shapes,arrows,automata,shadows,positioning,fit}
\usepackage{paralist} 
\usepackage{xspace}
\usepackage{alltt}
\usepackage{enumerate}
\usepackage[inline,shortlabels]{enumitem}
\usepackage{mathtools}
\usepackage{bm}
\usepackage{courier}
\usepackage{color}
\usepackage{caption}
\usepackage{stmaryrd}
\usepackage{adjustbox}
\usepackage{wrapfig}
\usepackage{todonotes}
\usepackage{pgfplots}
\usepackage{amstext} 
\usepackage{array}   
\usepackage[colorlinks=true,citecolor=blue,linkcolor=blue,urlcolor=blue]{hyperref}
\newcolumntype{L}{>{$}l<{$}} 
\usetikzlibrary{decorations.pathreplacing,calc}

\usepackage[algoruled,vlined,linesnumbered]{algorithm2e}
\SetKwInOut{Input}{input}
\SetKwInput{KwInput}{Input}                
\SetKwInput{KwOutput}{Output}              
\usetikzlibrary{decorations.pathreplacing,calc}

\newcommand{\CTL}{\textsf{CTL}}
\newcommand{\LTL}{\textsf{LTL}}
\newcommand{\LTLX}{\textsf{LTL$\setminus$X}}

\newcommand{\Sketch}{\textsf{Sketch}}
\newcommand{\PSketch}{\textsf{PSketch}}
\newcommand{\AlloyStar}{\textsf{Alloy}$^*$}

\newcommand{\NuSMV}{\textsf{NuSMV}}

\newcommand{\Party}{\textsf{Party}}

\newcommand{\X}{\bigcirc}
\newcommand{\Until}{\mathbf{U}}
\newcommand{\Always}{\Box}
\newcommand{\WUntil}{\mathbf{W}}
\newcommand{\Future}{\Diamond}
\newcommand{\CNF}{\mathsf{CNF}}
\newcommand{\NOT}{\mathsf{NOT}}
\newcommand\proj{\mathord{\uparrow}}

\newcommand{\traces}[1]{\mathit{Traces}(#1)}
\newcommand{\fairtraces}[1]{\mathit{FairTraces}(#1)}
\newcommand{\Path}[1]{\mathit{Path}(#1)}
\newcommand{\post}[2]{\mathit{Post}(#1,#2)}
\newcommand{\tcpost}[2]{\mathit{Post}^*(#1,#2)}
\newcommand{\lts}[1]{#1 = \langle S, \textit{Act}, \rightarrow, I, \textit{AP}, L \rangle}

\newcommand{\refin}[1]{\mathit{Ref}(#1)}

\newcommand{\lstfont}[1]{\color{#1}\scriptsize\ttfamily}

\newcounter{nalg} 
\renewcommand{\thenalg}{\arabic{nalg}} 
\DeclareCaptionLabelFormat{algocaption}{Algorithm \thenalg} 

\lstdefinestyle{Spec}{
    showstringspaces=false,
    backgroundcolor=\color{white},
    basicstyle=\lstfont{black},
    identifierstyle=\lstfont{black},
    keywordstyle=\color{black}\bfseries,
    numberstyle=\lstfont{black},
    stringstyle=\lstfont{cyan},
    commentstyle=\lstfont{red},
    emph={
        action, process,
        spec, invariant, main, 
    },
    emphstyle=\color{black}\bfseries,
    breaklines=true
}

\lstdefinestyle{Unity}{
    mathescape=true,
    showstringspaces=false,
    backgroundcolor=\color{white},
    basicstyle=\lstfont{black},
    identifierstyle=\lstfont{black},
    keywordstyle=\color{black}\bfseries\em,
    numberstyle=\lstfont{black},
    stringstyle=\lstfont{cyan},
    commentstyle=\lstfont{red},
    emph={
        var, Program, Process, begin, end, initial
    },
    emphstyle={\lstfont{black}\bfseries},
    breaklines=true
}

\lstnewenvironment{algo}[1][] 
{   
    \refstepcounter{nalg} 
    \setcounter{lstlisting}{\value{nalg}}
    \captionsetup{labelformat=algocaption,labelsep=colon} 
    \lstset{ 
        mathescape=true,
        frame=tB,
        numbers=left, 
        numberstyle=\tiny,
        basicstyle= \scriptsize\ttfamily,
        keywordstyle=\color{black}\bfseries\em,
        keywords={,input, output, return, datatype, function, in, if, else, foreach, while, begin, end, endif, endwhile, endfor, procedure, then, for, do, all, some, such, that,} 
        numbers=left,
        xleftmargin=.04\textwidth,
        emphstyle={\bfseries},
        morecomment=[l]{//},  
        #1 
    }
}
{}

\title{Bounded Synthesis of Synchronized Distributed Models from Lightweight Specifications} 
\titlerunning{Bounded Synthesis of Synchronized Distributed Models...}
\author{Pablo F.  Castro\inst{1,2},
Luciano Putruele\inst{1,2} \\
Renzo Degiovanni\inst{3},
Nazareno Aguirre\inst{1,2} 
}
\institute{Departamento de Computaci\'on, Universidad Nacional de R\'io Cuarto, Argentina \and
Consejo Nacional de Investigaciones Cient\'ificas y T\'ecnicas (CONICET), Argentina \and
Luxembourg Institute of Science and Technology, Luxembourg}

\begin{document}

\maketitle
\begin{abstract}
We present an approach to automatically synthesize synchronized  models from lightweight formal specifications.  Our approach takes as input a specification of a distributed system along with a global linear time constraint, which must be fulfilled by the interaction of the system's components.  It produces executable models for the component specifications (in the style of Promela language) whose concurrent execution satisfies the global constraint.  The component specifications consist of  a collection of actions described by means of pre and post conditions together with first-order relational formulas prescribing their behavior.  We use the \emph{Alloy Analyzer} to encode the component specifications and enumerate their potential implementations up to some bound,  whose concurrent composition is model checked against the global property.  Even though this approach is sound and complete up to the selected bound,  it is impractical as the number of candidate implementations grows exponentially.  
To address this, we propose an algorithm that uses batches of counterexamples to prune the solution space, it has two main phases:  \emph{exploration}, 
the algorithm collects a batch of counterexamples, and \emph{exploitation},   where this knowledge is used to speed up the search.
The approach is sound,  while  its completeness depends on the batches used.
We present a prototype tool,  describe some  experiments,  and compare it with  related approaches.

\end{abstract}

\section{Introduction} \label{sec:Intro}

Program synthesis \cite{MannaWolper84,PnueliRosner89,EmersonClarke82} allows for the automated construction of programs from specifications. Ideally, the user only provides a formal description of the intended behavior of a to-be-developed program, and then a program, correct with respect to the given specification, is automatically derived. A clear benefit of this ideal scenario  is that users only need to concentrate in writing software specifications, which are typically stated in a higher level of abstraction, that abstracts away involved implementation details. In the last decades, the constant advances in hardware, and the emergence of new synthesis methods, have made possible the application of synthesis to complex case studies (see, e.g.,  \cite{BGJPPW07a}). 

This paper focuses on the synthesis of high-level descriptions for distributed systems, such as protocols and multi-agent systems. These systems are crucial in modern software applications, including collaborative tools, communication networks, and cryptocurrencies. Moreover, their correct implementation is in general rather complex and time consuming,  which often results in subtle programming errors.
Particularly, when synchronization primitives (e.g., semaphores, locks, monitors) are used to manage access to shared resources.  Their incorrect use often leads to challenging issues like race conditions and deadlocks,  which are difficult to identify and fix using conventional methods. Thus, distributed algorithms, and in particular the use of synchronization primitives, constitute a compelling target for synthesis approaches.


Approaches to automated synthesis have seen significant improvement in recent years. Some techniques are based on game theory for temporal logic specifications \cite{Piterman+2006}, the synthesis from input-output examples \cite{Jha+2010}, inductive programming \cite{Abate+2018}, and techniques based on theorem proving \cite{Srivastava+2010}, which have been successfully applied in various contexts. Synthesis techniques for distributed algorithms are less common, as mentioned above,  in this setting we have  additional difficulties, such as the problem of dealing with several interacting processes, complying with specific communication architectures, and correctly instrumenting coordination via synchronization mechanisms. In fact, the problem of synthesizing distributed code from specifications is undecidable in many settings \cite{PnueliRosner90}.

The  (bounded) synthesis technique presented in this paper targets  asynchronous distributed systems that use shared memory for the communication and synchronize via locks.  A system specification, in this setting, is composed of a (finite) collection of process or component specifications consisting of actions specified via pre/post-conditions,  together with local constraints expressing required properties of the action  executions; and \emph{global} properties that the process/component interaction must guarantee.  Local constraints are expressed in Alloy's relational logic \cite{AlloyBook}, a first-order logic with transitive closure (necessary for expressing certain constraints, notably reachability), whereas global constraints are specified in linear-time logic.  If successful,  the technique produces implementations for the components that use their corresponding local actions as well as lock-based synchronizations, and satisfy the local constraints, while their interaction satisfies the global constraints.  

More specifically,  the approach encodes component specifications into an Alloy model \cite{AlloyBook}, in such a way that satisfying instances of the model correspond to locally correct implementations of the corresponding process. Thus,  we indirectly use Boolean satisfiability to enumerate (bounded) locally valid implementations of processes. When locally valid implementations of processes are found, their combination is checked against the \emph{global} specification, to verify whether the current local candidates lead to a correct distributed solution. For efficiency reasons, this latter global check is performed using a symbolic model checker. The search is performed in a lexicographic manner,  in which a backtracking is performed when the possible instances of a component specification are exhausted.
 However,  this basic algorithm is impractical  due to the exponential  number of potential component implementations,  for which we also have to consider all their possible combinations. 
 To address this, we introduce an improved algorithm that uses counterexample batches to speed up the process.  This algorithm  consists of two phases: an exploratory phase that collects instances from the solver and generates counterexamples via a model checker; followed by an exploitation phase that refines component specifications based on these counterexamples.  If no solutions are found, the process repeats with the refined specifications.  The algorithm's efficiency and completeness depend on the chosen batch configurations, which we explore in Section \ref{sec:examples}, discussing their benefits and drawbacks.
 
The paper is structured as follows.  In Section \ref{sec:background} we introduce preliminary concepts. Section \ref{sec:motivating-example} illustrates our approach via a motivating example.  Sections \ref{sec:algo} and \ref{sec:examples} describe the main algorithm and discuss experimental results. Finally,  Sections \ref{sec:related} and \ref{sec:conclusions} discuss related work and present some final remarks.


\section{Preliminaries} 
\label{sec:background}


\paragraph{Labeled Transition Systems.}

A well-established mechanism to provide semantics for reactive and distributed systems is via \emph{labeled transition systems} (LTSs). An LTS  is a tuple $\langle S, \textit{Act}, \rightarrow, I, \textit{AP}, L \rangle$ where: $S$ is a (finite) non-empty set of \emph{states}, $\textit{Act}$ is a (finite) set of \emph{actions}, $\rightarrow \subseteq S \times \textit{Act} \times S$ is a labeled transition relation, $I \subseteq S$ is a set of \emph{initial states}, $\textit{AP}$ is a (finite) set of atomic propositions, and $L:S \rightarrow 2^{\textit{AP}}$ is a labeling function. Intuitively, $S$ is an abstraction of the system states, $L$ indicates which atomic properties hold in each state, $\textit{Act}$ is an abstraction of the system actions, and $\rightarrow$ indicates how the system changes state as actions occur. For convenience, we write $s \xrightarrow{a} s'$ instead of $\langle s,a,s'\rangle \in \rightarrow$, and  $\post{s}{s'}$ if there is an $a \in \textit{Act}$ such that $s \xrightarrow{a} s'$. The reflexive-transitive closure of relation $\mathit{Post}$ helps capturing system executions, and is written $\mathit{Post}^*$.  A path over an LTS is a sequence of alternating states and actions $\pi=s_0, a_0,s_1, a_1 \dots$ such that $\forall i: s_i \xrightarrow{a_i} s_{i+1}$. We say that a path is an \emph{execution} (or \emph{trace}) if it is maximal. Without loss of generality, we assume that each state has at least one successor. When only states or transitions are relevant, we refer to executions as sequences of states or transitions. We extend $L$ to paths as follows: $L(\pi) = L(s_0, a_0,s_1, a_1,s_2, \dots) = L(s_0),L(s_1),L(s_2),\dots$. When convenient, we use notation $x(s)$ as an abbreviation of  $x \in L(s)$. An execution $\pi$ is \emph{fair} if $s,a,s'$ appears infinitely often in $\pi$ given that $s$ appears infinitely often in $\pi$ and $s \xrightarrow{a} s'$. From now on, $\Path{T}$ denotes  the set of paths of $T$, $\traces{T}$ denotes its set of executions, and $\fairtraces{T}$ denotes  its set of fair executions.

Throughout the paper we use the following notation, $(x_0,\dots,x_n){\uparrow}i$ denotes the $i$-th projection of tuple $(x_0,\dots,x_n)$. $[0,k]$ denotes the set $\{0,\dots,k\}$, $\prod_{i \in I} S_i$ refers to the Cartesian product of sets $S_i$, and $\coprod_{i \in I} S_i$ denotes the coproduct of sets $S_i$.

\paragraph{(Extended) First-Order Logic over LTSs.} Properties over LTSs will be expressed using first-order formulas. The logic is essentially the relational logic underlying Alloy \cite{AlloyBook}. For the sake of clarity, we will use standard first-order logic extended with transitive closure. Given actions $\textit{Act}$ and a set $\mathit{AP}$ of atomic propositions, the basic vocabulary consists of binary predicates $\{a \mid a \in \mathit{Act}\}\cup\{\mathit{Post}\}$, unary predicates $\{p \mid p \in \mathit{AP}\} \cup \{I\}$; the logical operators are the usual Boolean connectives and first-order quantifiers with variables ranging over states. We consider the reflexive-transitive closure operator, written $a^*$ (for any binary predicate $a$). Reflexive-transitive closure is necessary to express reachability constraints. Given an LTS $T$ and a formula $\phi$, we write $T \vDash \phi$ iff $\phi$ \emph{holds in} $T$. As a sample formula, $\exists s,s': I(s) \wedge \tcpost{s}{s'} \wedge p(s')$ states that $p$ holds in some state reachable from an initial state.

\paragraph{Linear-Time Temporal Logic.}	Linear-time temporal logic ({\LTL}) is a well-known logical formalism, that has been successfully used for specifying temporal properties of concurrent and reactive systems \cite{Katoen08}. Given a set $\textit{AP}$ of atomic propositions, the syntax of {\LTL} is inductively defined, using the standard logical and temporal operators, by the following grammar: $\Phi ::= p \mid \neg \Phi \mid \Phi \vee \Phi \mid  \X \Phi \mid \Phi \Until \Phi$, where $p \in \textit{AP}$. Additional operators, such as $\textit{true}$, $\textit{false}$, $\Future \phi$ (\emph{eventually} $\phi$), $\Always \phi$ (\emph{always} $\phi$) and $\phi \WUntil \psi$ ($\phi$ (weak) until $\psi$), can be defined as combinations of the above operators \cite{Katoen08}. {\LTL} formulas are interpreted over infinite sequences of Boolean assignments for variables in $\textit{AP}$, i.e., each sequence is of the form $\pi=\sigma_0 \sigma_1 \sigma_2 \dots \in (2^{\textit{AP}})^{\omega}$. The interpretation of {\LTL} formulas can be given as a satisfaction relation $\vDash$, defined recursively as follows: 
\begin{itemize}
\item $\sigma \vDash p$ iff $p \in \sigma_0$,
\item $\sigma \vDash \phi \vee \psi$ iff $\sigma \vDash \phi$ or $\sigma \vDash \psi$
\item $\sigma \vDash \X \phi$ iff $\sigma_1 \sigma_2 \dots \vDash \phi$ 
\item $\sigma \vDash \phi \Until \psi$ iff   $\exists k \geq 0: \sigma_k \sigma_{k+1} \dots \vDash \psi$ and $\forall 0 \leq i<k: \sigma_i \sigma_{i+1} \dots \vDash \phi$. 
\end{itemize}
{\LTLX} \cite{Katoen08} is the fragment of {\LTL} obtained by omitting operator $\X$ from formulas, typically used for expressing {\LTL} stutter invariant properties of concurrent systems \cite{PeledWilke1997}. We employ {\LTLX} to specify the global properties over distributed systems. Given a LTS $T$ and a {\LTL} formula $\phi$, we say that $T \vDash \phi$ (resp. $T \vDash_f \phi$) iff $L(\pi) \vDash \phi$ for every $\pi \in \traces{T}$ (resp. $\pi \in \fairtraces{T}$). 

\paragraph{Process Descriptions.}
\label{sec:programming-language} 
While transition systems serve as a low-level description of reactive system behavior, such systems are typically better described as a collection of interacting processes,  as done in languages like Promela.  For the sake of simplicity,  here we use a  simple guarded-command language, as those defined in  \cite{Dijkstra1975,AroraGouda93}.  A system will then consist of a collection of shared variables, and a collection of processes, that execute concurrently in an asynchronous way. Each process has its own collection of local variables, an initialization condition indicating in which (local) state each process starts execution, and a set of guarded actions that establish how a process can proceed execution. Each atomic action has the form $[a]B \rightarrow C$ where: $a$ is the name of the action,  $B$ (called the guard) is a Boolean statement over the (shared and local) variables of the program, and $C$ is a collection of assignments, written $x_0{:=}E_0, \dots, x_n{:=}E_n$, where the $x_i$'s  are local or shared variables, and $E_i$'s are expressions of the correct type. We assume that the test of $B$ and the assignments in $C$ are all executed atomically. The operational semantics of this process notation is simple. A state is an assignment of values to the variables in the program. An action $B \rightarrow C$ is enabled in a state, if $B$ is true in that state. An execution of $P$ is a sequence of states satisfying the following conditions: (i) the initial state satisfies the initial conditions, (ii) if state $s_i$ follows from $s_{i-1}$ in the sequence, then we have some action $B \rightarrow C$ such that $B$ holds at $s_{i-1}$ and executing $C$ results in $s_i$, (iii) the sequence is infinite (we assume that when no action is enabled then the last state is repeated). By this semantics, mapping process/system executions to LTSs is straightforward, and thus the definition of satisfaction of LTL properties by system/process executions is direct. 

\section{Motivating Example}\label{sec:motivating-example}
\sloppy We motivate and illustrate our technique with a well-known example of  a distributed system, Dijkstra's \emph{dining philosophers} \cite{Dijkstra71}.   There are $n$ philosophers sitting around a table and sharing $n$ forks.  Each philosopher has exactly two forks available, one to each side, shared with the adjacent  philosophers. On the table, there is a bowl of pasta, and the philosophers alternate between thinking and eating.  
\begin{wrapfigure}[17]{r}{.40\textwidth}
 \centering
 \vspace{-0.7cm}
 \includegraphics[scale=0.3]{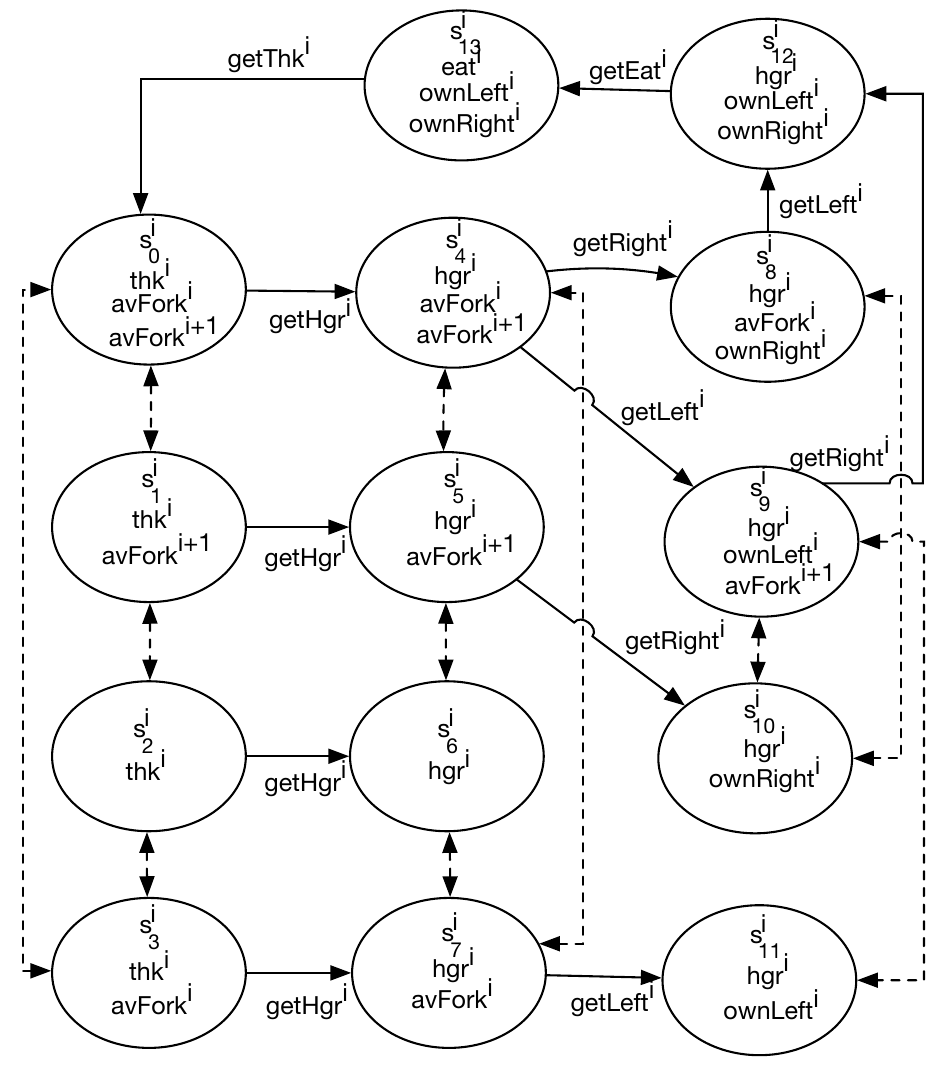} 
  \caption{LTS $T^i$ modelling philosopher $i$}\label{fig:philsEx}
\end{wrapfigure}
Each philosopher needs both forks to eat, and the access to a fork is (of course) mutually exclusive between its two users.  We want to design a  distributed protocol for this problem  that guarantees deadlock freedom. 
This problem admits many  solutions. For instance,  a way to avoid deadlock is implementing a policy where each philosopher picks up a fork only when her both forks are free.  Another possible solution is obtained  if  each philosopher first takes her left fork and then her right one,  excepting  one philosopher that takes the forks in the opposite order. These are distributed protocols, i.e., they do not depend on a centralised architecture. These are the kind of solutions we expect our technique to synthesize. 
\begin{wrapfigure}[25]{hr}{.25\textwidth}
    \vspace{-0.7cm}
    \includegraphics[scale=0.45]{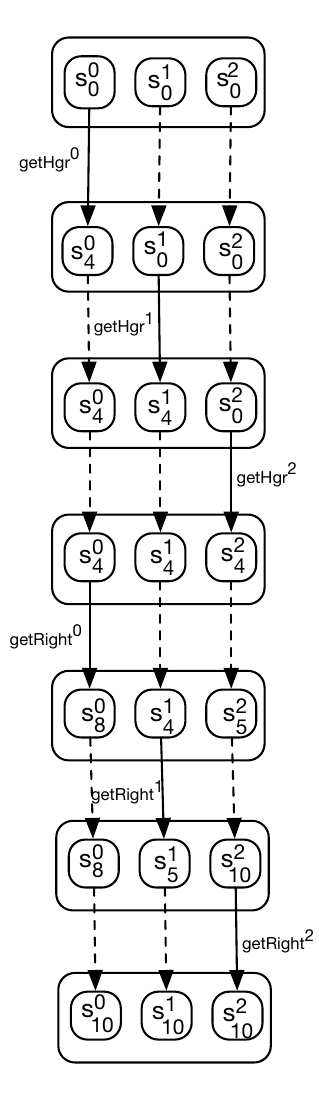} 
    \caption{A Path in $T^0 \parallel T^1 \parallel T^2$ leading to deadlock}\label{fig:philsTrace}
\end{wrapfigure}
Let us describe how this problem is specified in our setting.  The behavior of  philosopher's $i$ (for $0 \leq i < n$) is modeled  via the following set $\mathit{PS}^i$ of formulas:
\begin{enumerate}[(1)]
\item \label{phils-form-1} $\forall s : I(s) \Rightarrow \mathit{thk}^i(s) \wedge \neg \mathit{ownLeft}^i(s) \wedge \neg \mathit{ownRight}^i(s)$,
\item \label{phils-form-2}  $\forall s,s' : \neg \mathit{eat}^i(s) \Rightarrow \neg \textit{getThk}^i(s,s')$,
\item \label{phils-form-3} $\forall s,s' : \textit{getThk}^i(s,s') \wedge \mathit{eat}^i(s) \Rightarrow \mathit{thk}^i(s') \wedge \neg \mathit{ownRight}^i(s') \wedge \neg \mathit{ownLeft}^{i }(s') $,
\item  \label{phils-form-4} $\forall s,s': \textit{getHgr}^i(s,s') \Rightarrow \mathit{hgr}^i(s')$,
\item \label{phils-form-5} $\forall s, s':  \neg (\mathit{hgr}^i(s) \land \mathit{ownRight}^i(s) \land \mathit{ownLeft}^i(s)) \Rightarrow \neg \mathit{getEat}^i(s,s')$,
\item \label{phils-form-6} $\forall s,s': \textit{getEat}^i(s,s') \wedge \mathit{hgr}^i(s) \wedge \mathit{ownRight}^i(s) \wedge \mathit{ownLeft}^i(s) \Rightarrow \mathit{eat}^i(s')$,
\item \label{phils-form-7} $\forall s,s': \mathit{getRight}^i(s,s') \wedge \mathit{avFork}_{i+1}(s) \wedge \mathit{hgr}^i(s) \Rightarrow \mathit{ownRight}^i(s')$,
\item  \label{phils-form-8}$\forall s,s': \mathit{getLeft}^i(s,s') \wedge \mathit{avFork}_{i}(s) \wedge \mathit{hgr}^i(s) \Rightarrow \mathit{ownLeft}^i(s')$,
\item \label{phils-form-9}$\forall s \in S : I(s) \Rightarrow (\exists s' \in S: \tcpost{s}{s'} \wedge \mathit{eat}^i(s'))$,
\item \label{phils-form-10}$\forall s :  \mathit{avFork}_i(s)  \Rightarrow (\exists s' : \mathit{chFork_{i}}(s,s') \wedge \neg \mathit{avFork}_i(s'))$,
\item \label{phils-form-11}$\forall s :  \mathit{avFork}_{i+1}(s)  \Rightarrow (\exists s' : \mathit{chFork_{i+1}}(s,s') \wedge \neg \mathit{avFork}_{i+1}(s'))$.
\end{enumerate}

Thus, $\mathit{PS}^i$ describes the \emph{local} specification of philosopher $i$. These formulas may originate from a specification in a domain specific language, referring to actions (e.g., $\mathit{getEat}$), their corresponding guards (e.g., formula \ref{phils-form-5}) and effects (e.g., formula \ref{phils-form-6}). We do not deal with the design of a specification language in this paper, and directly refer to the set of formulas capturing the components behaviors. Notice how the local view of the system state is captured. Each philosopher uses Boolean variables $\mathit{ownLeft}^i$ and $\mathit{ownRight}^i$ for signaling the acquisition of the right and left fork, respectively. Variables $\mathit{avFork}_{i}$ and $\mathit{avFork}_{i+1}$ are used to indicate whether a fork is busy or not (being $+$ the addition modulo $n$).  The actions for philosopher $i$ are: $\mathit{getThk}^i$ (the philosopher goes to the thinking state), $\mathit{getHgr}^i$ (the philosopher gets hungry), $\mathit{getEat}^i$ (the philosopher goes to the eating state). Philosophers also have actions to obtain the forks: $\mathit{getRight}^i$ (obtain the right fork if available), and $\mathit{getLeft}^{i}$ (obtain the left fork of available). 
Variables $\mathit{avFork}_{i}$ and $\mathit{avFork}_{i+1}$ are shared with other philosophers. 
%
Formula \ref{phils-form-9} in $\mathit{PS}^i$ is a reachability (local) constraint, requiring that philosopher $i$ can  reach the eating state from the initial state. In this example, the forks act as locks: they are shared variables, and thus can be changed by the ``environment'' (from a local philosopher's perspective). This is captured through actions $\mathit{chFork}_{i}$ and $\mathit{chFork}_{i+1}$ that model the acquisition of the locks by the environment. Constraints \ref{phils-form-10}-\ref{phils-form-11} establish that, if a fork is free, it can be grabbed by another philosopher. 

The inputs for our technique include a \emph{global} temporal requirement, constraining the components interaction.  For instance,  in the case  $n=3$ we will have the local specifications $\mathit{PS}^0, \mathit{PS}^1$, and $\mathit{PS}^2$ together with the  {\LTL}  formula:
\begin{multline*}
   \Box \neg (\mathit{ownRight}^0 \wedge \mathit{ownRight}^1 \wedge \mathit{ownRight}^2) 
   \wedge \neg (\mathit{ownLeft}^0 \wedge \mathit{ownLeft}^1 \wedge \mathit{ownLeft}^2),
\end{multline*}
which states that  the system is deadlock-free.  

Note that, for each specification $\mathit{PS}^i$, one can employ a model finder to build an LTS $T^i$ that satisfies specification $\mathit{PS}^i$. Given the expressiveness of the logic, this can only be done up to a given bound on the size of the LTSs (the logic is undecidable). We use the Alloy Analyzer~\cite{AlloyBook}, a bounded model finder for relational logic (which subsumes first-order logic with transitive closure), for this task. Fig.~\ref{fig:philsEx} shows an LTS satisfying $\mathit{PS}^i$, obtained in this way. The dotted arrows represent environment transitions (which, once components are composed, will become actions performed by other philosophers). The global system behavior is obtained by the (asynchronous) parallel composition of the obtained LTSs, denoted  $T^0 \parallel T^1 \parallel T^2$. Now, we can verify whether the global system satisfies the global properties. In our example, if the LTSs $T^0$, $T^1$ and $T^2$ are as shown in Fig.~\ref{fig:philsEx}, then the global system will contain an execution leading to deadlock. 
The trace in Fig.~\ref{fig:philsTrace}, in which the three philosophers get hungry in turns, and then take their corresponding right fork, leads to deadlock. 

Our approach exploits counterexamples to global properties to guide the search for local implementations.  For instance,  we force at least one of the local implementations to avoid the consecutive execution of $\textit{getHgr}^i$ and $\mathit{getRight}^i$. Assuming that under this new constraint we obtain an LTS $T'^{0}$ for philosopher $0$, that takes the left fork before taking the right one, then the global system $T'^{0} \parallel T^1 \parallel T^2$ is deadlock-free and satisfies the global temporal requirement.


\section{System Specifications}

In this section, we provide a more formal and detailed definition of specifications, as introduced in the previous section.  In the following definitions we  mainly use the first-order logic with transitive closure described in Section~2.  We start by giving a precise definition of the notion of component specification.

\begin{definition} A \emph{process (or component) specification} $\mathit{PS}$ is a tuple $\langle \langle \mathit{Sh},  \mathit{Loc},$ $\mathit{Act}\rangle, \Phi \rangle$ where $\mathit{Sh}$, $\mathit{Loc}$, $\mathit{Act}$ are finite and mutually disjoint sets, and $\Phi$ is a finite set of first-order (relational) formulas over the vocabulary defined by $\mathit{Sh} \cup \mathit{Loc} \cup \mathit{Act}$.
\end{definition}
Intuitively,  \textit{Sh} are the shared variables used by the process,  \textit{Loc} are the process's local variables, and \textit{Act} are the process's actions. A process specification defines a collection of LTSs satisfying the requirements in the specification. Given a process specification $\textit{PS}=\langle \langle \mathit{Sh},  \mathit{Loc}, \mathit{Act}\rangle, \Phi \rangle$ and an LTS $T=\langle S, \mathit{Act}, \rightarrow, I, \mathit{Sh}\cup\mathit{Loc} , L\rangle$, we write $T \vDash PS$ iff $T \vDash \phi$ for every $\phi \in \Phi$.
 
A specification of a distributed system is a collection of process specifications with the same shared variables, and an additional global temporal requirement.
\begin{definition} A system \emph{specification} $\mathcal{S}$ is a tuple $\langle \{ \textit{PS}^i \}_{i \in \mathcal{I}},$ $ \phi \rangle$, where $\mathcal{I}$ is a finite index set, each $\textit{PS}^i = \langle \langle \mathit{Sh}, \mathit{Loc}^i, \mathit{Act}^i\rangle,$ $\Phi^i \rangle$ is a \emph{local} process specification, and $\phi$ is a \emph{global} requirement expressed by an {\LTL} formula over the vocabulary $\mathit{Sh} \cup \coprod_{i \in \mathcal{I}} \mathit{Loc}^i$.
\end{definition}

By simply considering that a system specification $\mathcal{S}$ is a collection of component specifications, we do not provide any specific semantics to shared variables, compared to process local variables. As a consequence,  locally correct implementations may often lead to invalid system implementations (system implementations violating the constraints), since when putting the process implementations together,  any local assumptions on shared variables may simply not hold. This can be solved by forcing local process specifications to assume unrestricted behavior of the environment, which is formalized as follows: 
 
 \begin{definition}\label{def:system-spec}
Let $\mathcal{L} = \{\ell_0, \dots, \ell_m \}$ be a set of elements (called \emph{locks}). An $\mathcal{L}$-\emph{synchronized specification} is a system specification $\mathcal{S} = \langle \{\textit{PS}^i\}_{i \in \mathcal{I}}, \phi \rangle$ with \emph{shared} variables $\{\mathit{av}_{\ell_0}, \dots,$ $\mathit{av}_{\ell_m}\}  \subseteq \mathit{Sh}$, and where each process specification $\textit{PS}^i = \langle \langle \mathit{Sh}, \mathit{Loc}^i, \mathit{Act}^i\rangle, \Phi^i \rangle$ has local variables  $\{ \mathit{own}_{\ell_0}, \dots,$ $\mathit{own}_{\ell_m} \} \subseteq \mathit{Loc}^i$, and actions $\{\textit{ch}_{\ell_0},\dots,\textit{ch}_{\ell_m}\} \cup \{\mathit{ch}_g \mid g \in \mathit{Sh}\} \subseteq \mathit{Act}^i$. Furthermore, the following formulas belong to each $\Phi^i$:
\begin{enumerate}[(a)]
	\item\label{system-spec-formula1}	 $\bigwedge_{\ell \in \mathcal{L}}(\forall s : \textit{own}_\ell(s) \Rightarrow   \neg \mathit{av}_\ell(s))$,
	\item\label{system-spec-formula2} $\bigwedge_{\ell \in \mathcal{L}}(\forall s  : \neg \textit{own}_\ell(s) \equiv (\exists s'  : s \overset{\textit{ch}_\ell}{\rightarrow} s'))$,
	\item\label{system-spec-formula3} $\bigwedge_{\ell \in \mathcal{L}}(\forall s,s'  : s \overset{\textit{ch}_\ell}{\rightarrow}{s'} \Rightarrow (\textit{av}_\ell(s) \equiv  \neg \textit{av}_\ell(s'))$,
	\item\label{system-spec-formula4} $\bigwedge_{\ell \in \mathcal{L}}\bigwedge_{v  \in (\textit{Sh}\cup\mathit{Loc} \setminus \{\mathit{own}_\ell, \mathit{av}_\ell\})}(\forall  s,s'  : s \overset{\textit{ch}_\ell}{\rightarrow} s' \Rightarrow (v(s) \equiv v(s'))$,
	\item\label{system-spec-formula5}   $\bigwedge_{g \in \textit{Sh}\setminus\{\mathit{av}_{\ell_0}, \dots,\mathit{av}_{\ell_m}\} }(\forall s: (\exists s': s \overset{ch_g}{\rightarrow} s' \wedge g(s')) \wedge  (\exists s': s \overset{ch_g}{\rightarrow} s' \wedge \neg g(s')))$,
	\item\label{system-spec-formula6} 	$\bigwedge_{g \in \mathit{Sh}}\bigwedge_{g'  \in (\textit{Sh}\cup\mathit{Loc} \setminus \{g\})}(\forall  s,s'  : s \overset{\textit{ch}_g}{\rightarrow} s' \Rightarrow (g'(s) \equiv g'(s'))$.
\end{enumerate}
\end{definition}
Intuitively, locks are special kinds of shared variables used for synchronization.  Actions $\{\textit{ch}_{\ell_0},\dots,\textit{ch}_{\ell_m}\} \cup \{\mathit{ch}_g \mid g \in \mathit{Sh}\}$ 
are used to model environment actions, for instance, $\textit{ch}_{\ell_0}$ represents an action of the environment that changes the state of lock $\ell_0$.  Variables $\mathit{av}_i$ and
$\mathit{own}_i$ are used to indicate that a lock is free,  or owned by the current component.  Formula \ref{system-spec-formula1} expresses that, if a component owns a lock, then it is not available.
Formula \ref{system-spec-formula2}  expresses that, if a lock is not owned by the current component, then it can be changed by the environment.  Formula \ref{system-spec-formula3} says that, if the environment changes lock $\ell_i$, then the variable 
$\mathit{av}_i$ changes accordingly.  Formula \ref{system-spec-formula4} states that changes in the locks do not affect other variables.  Formula \ref{system-spec-formula5} expresses that shared variables can be changed by the environment. Finally, formula \ref{system-spec-formula6} states that the changes in one shared variable does not affect the other variables.
From now on, we write $s \dashrightarrow s'$ if $s \xrightarrow{\textit{ch}_\ell} s'$ or $s \xrightarrow{\textit{ch}_g} s'$, for some lock $\ell$ or shared variable $g$, respectively.

In the same way as component specifications can be put together, we also consider the asynchronous composition of LTSs.
\begin{definition} Let $\mathcal{S} = \langle \{\textit{PS}^i\}_{i \in [0,n]}, \phi \rangle$ be an $\mathcal{L}$-\emph{synchronized specification}, and 
let $T^0,\dots,T^n$ be LTSs, such that for each $T^i = \langle \mathit{S}^i, \mathit{Act}^i, \rightarrow^i,I^i, \mathit{Sh}\cup \mathit{Loc}^i, L^i \rangle$ we have $\mathit{Sh}\cap \mathit{Loc}^i = \emptyset$ and $T^i \vDash \mathit{PS}^i$. We define the LTS $T^0 \parallel \dots \parallel T^n = \langle S, \coprod_{i \in [0,n]} \mathit{Act}^i, \rightarrow, I, \mathit{Sh}\cup \coprod_{i \in [0,n]}\mathit{Loc}^i, L\rangle$   as follows:
\begin{itemize}	
	\item $S = \{s \mid s \in \prod_{i \in [0,n]}S^i \wedge (\forall g \in \mathit{Sh} : \forall i,j \in [0,n] : L^i(s \proj i)(g) = L^j(s \proj j)(g)) \}$,
	\item $\rightarrow  = \{ s \xrightarrow{a} s' \mid \exists i \in [0,n] : (s \proj i \xrightarrow{a}\mathrel{\vphantom{\to}^i}  s' \proj i) \wedge (\forall j \neq i : (s \proj j \dashrightarrow^j s' \proj j) \vee (s \proj j  =  s' \proj j) \}$,
	\item $I  =\{ s \mid \forall i \in [0,n]: s{\uparrow}i \in I^i\}$,	
	\item  $L(s)(x)=  \begin{cases*}
      					L^i(s \proj i)(x) & if  $x \in Loc^i$ for some $i \in [0,n]$, \\
     					L^0(s \proj 0)(x)      & otherwise.
   				  \end{cases*}$
\end{itemize}
\end{definition}
The semantics of distributed specifications is straightforwardly defined using asynchronous product, i.e., the combination of LTSs that produces all interleavings of their corresponding actions.
\begin{definition} Given a system specification $\mathcal{S} = \langle \{\mathit{PS}^i\}_{i \in [0,n]},$ $\phi \rangle$ a \emph{model} or \emph{implementation} of $\mathcal{S}$ is a collection of LTSs $T^0,\dots,T^n$ such that $T^i \vDash \mathit{PS}^i$, for every $i \in [0,n]$, and $T^0 \parallel \dots \parallel T^n \vDash \phi$.
\end{definition}
An interesting point about the definition above, which in particular holds thanks to the formulas that give semantics to shared variables, is that linear-time temporal properties (without the next operator) local to the process implementations can be promoted to global implementations, i.e., to asynchronous products they participate in, provided the asynchronous product is \emph{strongly fair} (strong fairness assumption is necessary to guarantee the promotion of liveness properties). That is, LTSs satisfying \ref{system-spec-formula1}-\ref{system-spec-formula6} preserve their (local) temporal properties under \emph{any} environment that guarantees strong fairness:
\begin{theorem}\label{theorem:stuttering-equiv} Let $\langle \{ \mathit{PS}^i \}_{i \in [0,n]}, \phi \rangle$ be a distributed specification, and $T^0, \dots, T^n$ LTSs such that $T^i \vDash \mathit{PS}^i$ (for every $i \in [0,n]$). Given an {\LTLX} formula $\psi$, if $T^i \vDash \psi$  (for any $i \in [0,n]$), then $T^0 \parallel \dots \parallel T^n \vDash_f \psi$.
\end{theorem}
	
From an LTS $T$, we can obtain a specification that characterizes it (up to isomorphism), using existentially quantified variables for identifying the states, and formulas for describing the transitions. Moreover,we can obtain a specification that characterizes \emph{all} LTSs that can be obtained by removing some (local) transitions from $T$. Intuitively,  this specification captures refinements of $T$. 
\begin{definition}\label{def:ref-spec} Let $\mathit{PS}= \langle \langle \mathit{Sh},  \mathit{Loc}, \mathit{Act}\rangle, \Phi \rangle$ be a component specification, and let $T=\langle S, \mathit{Act}, \rightarrow, I, (\mathit{Sh} \cup \mathit{Loc}), L\rangle$ be an LTS, such that $S=\{s_0,\dots,s_n\}$, $\mathit{Sh} \cup \mathit{Loc}=\{p_0,\dots,p_m\}$, $\mathit{Act}=\{a_0,\dots,a_k\}$ and $T \vDash \mathit{PS}$. The process specification $\refin{\mathit{PS},T}$ is the tuple $\langle \langle \mathit{Sh}, \mathit{Loc}, \mathit{Act}\rangle, \Phi \cup \{ \exists s_0,\dots,s_n: \phi^T\} \rangle$ where $\phi^T$ is the following formula:
 \[\displaystyle
\begin{array}{l}(\bigwedge_{0\leq i< j \leq n} s_i \neq s_j) \wedge I(s_0) 
															\wedge \bigwedge_{j \in [0,n]} \bigwedge_{i \in [0,m]} \{  p_i(s_j) \mid p_i \in L(s_j) \}  \\
															\wedge  \bigwedge_{j \in [0,n]} \bigwedge_{i \in [0,m]} \{ \neg p_i(s_j) \mid p_i \notin L(s_j) \} \\
															\wedge \bigwedge_{j,j' \in [0,n]} \bigwedge_{i \in [0,k]}  \{ \neg a_i(s_j,s_{j'}) \mid \neg (s_j \xrightarrow{a_i} s_{j'}) \}\\
															\wedge \bigwedge_{j,j'\in [0,n]} \bigwedge_{i \in [0,k]} \{ a_i(s_j,s_{j'}) \mid 
															s_j \xrightarrow{a_i},s_{j'} \wedge a_i \in \{\textit{ch}_g \mid g \in \mathit{Sh} \} \}   
																													
\end{array}				
\]
\end{definition}
Formula $\exists s_0,\dots,s_n: \phi^T$ identifies each state with a variable, describes the properties of each state, enumerates the transitions labeled with environment actions, and rules out the introduction of local transitions not present in $T$. Summing up, models of $\refin{\mathit{PS}, T}$ may remove some local transitions present in $T$ but still satisfy specification $\mathit{PS}$. It is direct to see that $T \vDash \refin{\mathit{PS},T}$. Furthermore, $\refin{\mathit{PS},T}$ preserves all the safety properties of $T$ (cf. \cite{Katoen08} for a formal definition of safety).
\begin{theorem}\label{theorem:preserve-safety} Let $\phi$ be an {\LTL} safety formula, $\mathit{PS}$ a process specification, and $T$ an LTS such that $T \vDash \mathit{PS}$. Then:
$
	T \vDash \phi \text{ implies } \refin{\mathit{PS},T} \vDash \phi. 
$
\end{theorem}

The following notation will be useful in later sections, to refer to specifications complemented with additional formulas.


\begin{definition} Let $\mathit{PS}= \langle \langle \mathit{Sh},  \mathit{Loc}, \mathit{Act}\rangle, \Phi \rangle$ be a process specification, and $T=\langle S, \mathit{Act}, \rightarrow, I, (\mathit{Sh} \cup \mathit{Loc}), L\rangle$ an LTS such that $S=\{s_0,\dots,s_n\}$, and $T \vDash \mathit{PS}$. Given a formula $\psi$ with free variables $s_0,\dots,s_n$, we define:
\[
	\refin{\mathit{PS},T} \oplus \psi = \langle \langle \mathit{Sh}, \mathit{Loc}, \mathit{Act}\rangle, \Phi \cup \{ \exists s_0,\dots,s_n: \phi^T \wedge \psi\} \rangle
\]	
\end{definition}
\section{Obtaining Guarded-Command Programs.}
    In this section we describe how we obtain the guarded-command programs from LTSs.  It is worth noting that the programs we synthesize  use locks for achieving synchronization. When a guard checks for a lock availability and this is not available  the process may continue executing other branches (i.e., our locks are not blocking). However, note that a process could get blocked when all its guards are false, thus other synchronization mechanisms such as blocking locks, semaphores and condition variables can be expressed by these programs. 
    
	Given LTSs $T^i = \langle \mathit{S}^i, \mathit{Act}^i, \rightarrow^i,I^i, \mathit{Sh}\cup \mathit{Loc}^i, L^i \rangle$ for $i \in [0,n]$, we can abstract away the environmental transitions and define a corresponding program in  guarded-command notation,  denoted $\text{Prog}(T^0  \parallel \dots \parallel T^{n})$, as follows. The shared variables are those in $\textit{Sh}$ plus an additional shared variable $\ell$ for each lock, with domain $[0,n-1]\cup\{\bot\}$ (where $\bot$ is a value indicating that the lock is free). Additionally, for each $T^i$ we define a corresponding process. To do so, we introduce for each $s \in S^i$  the equivalence class: $[ s ]  = \{ s' \in S^i \mid (s \dashrightarrow^* s') \vee (s'  \dashrightarrow^* s)\}$. 
That is, it is the set of states connected to $s$ via environmental transitions. It is direct to see that it is already an equivalence class. The collection of all equivalence classes
is denoted $S^i /_{\dashrightarrow^*}$.The local variables of the process are those in $\textit{Loc}^i$ plus a fresh variable $\textit{st}_i$, with domain $S^i/_{\dashrightarrow^*}$ (for indicating the current state of the process). 

Finally, given states $s,s' \in S^i$ with $s \xrightarrow{a} s' \in \rightarrow^i$ and  $[s] \neq [s']$, we consider the following  guarded command:
\[
 [a] \left( \begin{array}{l} 
 			 state = [s] \\
			 \wedge \bigwedge \{x = x(s) \mid x \in \mathit{Sh} \cup \mathit{Loc} \}\\ 
			\wedge \bigwedge \{ \ell = i \mid \ell \in \mathcal{L} : \mathit{own}_\ell \in L(s)\}	 \\ 
			 \wedge \bigwedge \{ \ell = \bot \mid \ell \in \mathcal{L} : \mathit{av}_\ell \in L(s)\}  \end{array} \right) \rightarrow \left( \begin{array}{l} 
 																					  			  \{x {:=} x(s')  \mid x \in \mathit{Loc}\cup \mathit{Sh}\} \\ 
																					 			  \cup \{\mathit{state} {:=} [s'] \} \\
																								  \cup \{ \ell := i \mid \ell \in \mathcal{L} : \mathit{own}_\ell \in L(s)\}  \\
																								  \cup \{ \ell := \bot \mid \ell \in \mathcal{L} : \mathit{av}_\ell \in L(s)\}
																					  \end{array} \right)
\]

We can prove that our translation from transition systems to programs is correct. That is, the executions of the program satisfy the same temporal properties as the 
asynchronous product $T^0 \parallel \dots \parallel T^{n}$.
\begin{theorem}\label{th:proppreservation} Given LTSs $T^i$ and \textsf{LTL} property $\phi$, then we have that:
$
	T^0 \parallel \dots \parallel T^{n} \vDash \phi \Leftrightarrow Prog(T^0 \parallel \dots \parallel T^{n}) \vDash \phi
$
\end{theorem}
\begin{figure}[t!]
\begin{minipage}[b]{0.30\linewidth}
\centering
\includegraphics[scale=0.6]{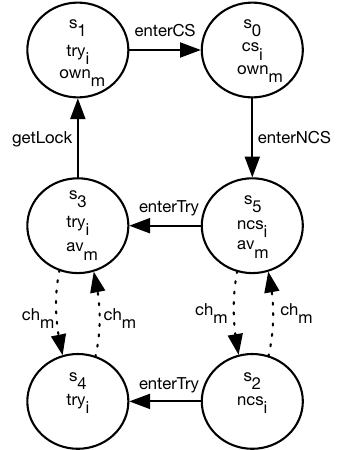}\label{fig:mutex-lts}
\end{minipage}
\hspace{0.7cm}
\begin{minipage}[b]{0.70\linewidth}
\centering
\begin{lstlisting}[style=Unity]
Program Mutex
 var m:Lock;
  Process $P_i$ with $i \in \{0,1\}$
   var $\text{try}_i, \text{ncs}_i, \text{cs}_i$:bit
   var $\text{st}_i$:$\{\text{S0},\text{S1},\text{S2},\text{S3},\text{S4},\text{S5}\}$
   initial: $\text{ncs}_i\wedge \neg \text{cs}_i \wedge \neg \text{try}_i$ 
   begin
    [enterTry]$\text{st}_i$=S5 $\rightarrow$ $\text{st}_i$:=S3,$\text{try}_i$:=1,$\text{ncs}_i$:=0
    [getLock]$\text{st}_i$=S3$\wedge$ m=$\bot$ $\rightarrow$ $\text{st}_i$:=S1,$\text{try}_i$:=1,m:=i
    [enterCS] $\text{st}_i$=S1$\rightarrow$ $\text{st}_i$:=S0,$\text{cs}_i$:=1,$\text{try}_i$:=0
    [enterNCS]$\text{st}_i$=S0$\rightarrow$st:=S5,$\text{ncs}_i$:=1,$\text{cs}_i$:=0 m:=$\bot$
   end
end
\end{lstlisting}
\end{minipage}
\caption{LTS and corresponding program for mutex}\label{fig:mutex}
\end{figure}
\begin{example}[Mutex]  
\label{ex:mutex-ts}
Consider a system composed of two processes (it can straightforwardly be generalized to $n$ processes) both with non-critical, waiting and critical sections. The global property  is mutual exclusion: the two processes cannot be in their critical sections simultaneously. We consider one lock $m$,  actions: $\mathit{enterNCS}$ (the process enters to the non-critical section), $\mathit{enterCS}$ (the process enters to the critical section), $\mathit{getLock}$ (the process acquires the lock), $\mathit{enterTry}$ (the process goes to the try state), and the corresponding propositions $\mathit{ncs}, \mathit{cs}, \mathit{try}, \mathit{own}_m, \mathit{av}_m$.  

The transition system and the program corresponding to this example are shown in Fig.~\ref{fig:mutex}. It is interesting to observe that,  the  conditions in Definition \ref{def:system-spec} allow one to use the locks as synchronization mechanisms, if the component $i$ owns the lock $m$, then the other processes cannot go into their critical sections since the lock will be not available for them. Also,  it is worth noting that we consider one state in the program for each equivalence class of states in the LTS.
\end{example}

\vspace{-0.5cm}


\section{Synthesis Algorithm}\label{sec:algo}
In this section, we detail the algorithm we propose for synthesizing synchronized process models from specifications.
This algorithm takes as input a specification $\mathcal{S} = \langle \{ \textit{PS}^i \}_{i \in \mathcal{I}}, \phi \rangle$ for a distributed system, and attempts to synthesize  local implementations  $\{ T^i \}_{i \in \mathcal{I}}$, that satisfy the corresponding local specification (i.e., $T^i \models \mathit{PS}^i$, for every $i \in \mathcal{I}$); and whose  asynchronous parallel composition satisfies the global temporal requirement $\phi$, i.e., $\parallel_{i \in \mathcal{I}} T^i \models \phi$. The algorithm also uses a bound  $k$ to limit the maximum number of states for the synthesized LTSs.  When an implementation is found, it is returned in the {\NuSMV} language; if not, the algorithm deems the specification as unsatisfiable within the bound $k$. 

First, we present Alg.~\ref{alg:the_alg},  a basic  version of the synthesis procedure. 
In line \ref{alg1-for-loop}, it inspects all instances of the current specification.  Line \ref{alg1-base-case} handles the base case (the last specification): a model checker is called to verify the selected instances; if successful the obtained solution is returned; otherwise, another instance is chosen. Line \ref{alg1-recursive-call} addresses the recursive case, where a different instance is examined, and the algorithm continues recursively with the next specification.
As shown in Section \ref{sec:examples}  this algorithm can only cope with small problems. 
{\scriptsize
\begin{algorithm*}[t]
\caption{A Simple Search Algorithm}
\label{alg:the_alg}
\SetKwProg{Fn}{Function}{}{end}
\SetKwFunction{Synt}{Synt}%
\SetFuncSty{textbf} 
\KwData{$i$ (process number)
$\textit{PS}^0,\dots,\textit{PS}^n$ (process specifications), $\phi$ (global property), $k$ (instance bound)}
\Fn{\Synt{$i,\textit{PS}^0,\dots,\textit{PS}^n, k $}}{
\KwResult{$r_0, \dots, r_n$ with $r_i \vDash \textit{PS}^i$ and
$r_0 \parallel \dots \parallel r_n \vDash \phi$ or $\emptyset$ otherwise}
 \For{all instances $M_i$ of $\textit{PS}^i$}{  \label{alg1-for-loop}
            $r_i \gets M_i$\;
            \lIf{$i=n \wedge r_0 \parallel \dots \parallel r_n \vDash \phi$}{ \Return{$\{r_0,\dots,r_n $\}} } \label{alg1-base-case}
             \If{$i < n$}{
                $\textit{found} \gets \Synt(i+1,\textit{PS}^0, \dots, \textit{PS}^n, k)$\; \label{alg1-recursive-call}
                \lIf{$\textit{found} \neq \emptyset$}{\Return{$\textit{found}$}}
            }
   }
\Return{$\emptyset$}\;
}
\end{algorithm*}
}
One main issue with Alg.~\ref{alg:the_alg} is that a wrong choice for the initial instance of the first specifications  (in lexicographic order, e.g., $\textit{PS}^0$) could result in the algorithm
getting stuck while searching for instances of the last specifications (e.g.,  $\textit{PS}^n$) before coming back to the initial specifications to try with different instances of them.  

We improve Alg.~\ref{alg:the_alg} using two main techniques: first, we force the synthesizer to start from better instances (see below); second, we use  counterexamples to speed up the search.
For the first point we add to the specifications formulas that ensure that the obtained instances contain a large number of transitions, which can be pruned in later stages.  Specifically, for every action $\textit{act}$ in a specification $\textit{PS}^i$ we add the formula:
\[
    \forall s \in S : \textit{Pre}_{\textit{act}}(s) \Rightarrow \exists s' \in S : \textit{act}(s,s')
\]
which states that, for every state where the precondition of the action ($\textit{Pre}_{\textit{act}}$) holds,  an execution of $\textit{act}$ can be observed.  This ensures that the obtained model has a good amount of transitions.  We use this for the second technique we apply,  namely, we use counterexamples to prune action executions that may violate the global formulas.  This approach is obviously sound but not complete, because a wrong selection of initial instances may imply that no solution will be found. However, in Section \ref{sec:examples} we show that this procedure is able to deal with an interesting set of examples.  
Furthermore, to avoid that the algorithm gets stuck while inspecting instances of the last specifications without comebacking to try new instances of the first components, we introduce the use of batches.
More specifically,  we use a sequence of bounds $b_0 \dots b_n$
aimed at performing a bounded backtracking. Furthermore,   the counterexamples collected in each batch are utilized for speeding up the process.  Roughly speaking,  we first execute the algorithm inspecting only $b_0$ instances of each specification; if no solution is found,  we use the counterexamples collected  but now with $b_1$ batches,  and so on. 

Some words are useful about the way in which our algorithm uses counterexamples for improving the search. Given the LTSs $\{ T^i \}_{i \in \mathcal{I}}$ and the global {\LTLX} property $\phi$, a \emph{counterexample} of $\parallel_{i \in \mathcal{I}} T^i \models \phi$ is an execution $\pi$ of $\parallel_{i \in \mathcal{I}} T^i$ such that $\pi \nvDash \phi$.  Our modified algorithm takes a counterexample $\pi$ generated by a model checker, and \emph{projects} this global execution to local executions of the participating processes. This information is then used to \emph{refine} the local process specifications to get rid of the projected counterexamples, and explore new implementations for the processes participating in the removed counterexample.

In finite transition structures, counterexamples can be represented by finite paths, called \emph{lasso traces} (a finite sequence of states, such that the last state has a loop to some previous state) \cite{Biere+1999}.  Noting that any finite path can be identified with a formula in \emph{conjunctive normal form} yields the following definition.

\begin{definition}Given an LTS $\lts{T}$ and a finite path $\pi = s_0, s_1, \dots ,s_{m} \in \Path{T}$, we define the formula:
$
\CNF(\pi) = \bigwedge_{0 \leq j < m}(\bigvee_{a \in \mathit{Act}}(a(s_j,$ $s_{j+1})),
$
where $s_0,\dots,s_m$ are free variables.
\end{definition}

The idea behind this definition is that paths can be captured by means of formulas. Let us denote by  $\llbracket \CNF(\pi) \rrbracket$ the set of clauses of formula $\CNF(\pi)$. For instance, $\refin{\mathit{PS},T} \oplus \CNF(\pi)$ captures the refinements of $T$ satisfying specification $\mathit{PS}$ and preserving path $\pi$. Similarly, we can define a formula that removes a counterexample from the instances of a specification.

\begin{definition}\label{def:not} Let $\lts{T}$ be an LTS and $\pi = s_0, s_1, \dots, s_m \in \Path{T}$ a path such that $s_i \neq s_{i+1}$ for some $i$. We define the following formula over $T$:
$
\NOT(\pi) = \bigvee_{0 \leq i < m} (\bigwedge_{a \in \mathit{Act}} \neg  a(s_{i},s_{i+1})  \wedge \ (s_i \neq s_{i+1})),
$
where $s_0,\dots,s_m$ are free variables.
\end{definition}

$\refin{\mathit{PS},T} \oplus \NOT(\pi)$ identifies the refinements of $T$ that do not have path $\pi$, as proved by the following theorem.

\begin{theorem}\label{theorem:cex-entails} Let $\mathit{PS}$ and $T$ be a specification and an LTS, respectively, such that $T \vDash \mathit{PS}$, and $\pi = s_0,\dots, s_m \in \Path{T}$ , with $s_i \neq s_{i+1}$ for some $i$. Then:
$
	\refin{\mathit{PS},T} \oplus \NOT(\pi)  \nvDash \CNF(\pi).
$
\end{theorem}
	
Given LTSs $\{ T^i \}_{i \in \mathcal{I}}$ and a finite path $\pi = s_0, \dots, s_m \in \Path{\parallel_{i \in \mathcal{I}} T^i}$, we denote by $\pi {\uparrow} i$ the path projected to an execution of process $T^i$, defined as  $\pi {\uparrow} i = s_0 {\uparrow} i, \dots, s_m {\uparrow} i$. Projecting a global execution may introduce stuttering steps in local executions. The projections of a global execution form a tuple  $\langle \pi{\uparrow}i \rangle_{i \in \mathcal{I}}$, that will be used to refine the instances obtained via the satisfiability solver. 

Using these definitions we introduce our updated algorithm.  Alg.~\ref{alg:improved_alg} implements the ideas previously discussed.  The Algorithm consists of a starting function (\textbf{StartSearch}) that sets the starting models of each specification (line \ref{alg2-initial-models}), as explained above these initial models are obtained by enriching the specification with specific formulas.  Line \ref{alg2-counterexamples-init}
initializes each specification with a tailored specifications $\textit{Ref}(\textit{PS}^i,M_i) \oplus \bigwedge_{\pi \in \textit{cexs}} \NOT(\pi{\uparrow}i)$.  Intuitively, these specifications consider all the refinements of the initial LTSs together with a set of counterexamples that can be used for refining them.  Line \ref{alg2-call-batchsynth} calls to the auxiliary function \textbf{BatchSynt}, which has similar behavior to Alg.~\ref{alg:the_alg}, but it takes into account the bounds for each batch  (line \ref{alg2-batch-loop}).
{\scriptsize
\begin{algorithm*}[t]
\caption{Batches Algorithm}
\label{alg:improved_alg}
\SetKwComment{Comment}{/* }{ */}
\SetKwProg{Fn}{Function}{}{end}
\SetKwFunction{StartSearch}{StartSearch}%
\SetKwFunction{BatchSynt}{BatchSynt}%
\SetFuncSty{textbf} 
\KwData{$\textit{PS}^0,\dots,\textit{PS}^n$ (process specifications), $\phi$ (global property), $k$ (instance size bound), $b_0,\dots,b_m$ (sequence of batch bounds), $cexs$ (global set of counterexamples)}
\Fn{\StartSearch{$\textit{PS}^0,\dots,\textit{PS}^n$, k,  $b_0,\dots,b_m$}}{
\KwResult{$r_0, \dots, r_n$ with $r_i \vDash \textit{PS}^i$ and
$r_0 \parallel \dots \parallel r_n \vDash \phi$ or $\emptyset$ otherwise}
$cexs \gets \emptyset$\;
 \lForEach{ i = 0 \dots n}{$M_i \gets \text{initial instance of } \textit{PS}^i \textit{of size } k$} \label{alg2-initial-models}
\For{$b = b_0,\dots,b_m$}{ \label{alg2-batch-loop}
    \lForEach{$i=0\dots n$}{  $\textit{PS}^i \gets \textit{Ref}(\textit{PS}^i,M_i) \oplus \bigwedge_{\pi \in \textit{cexs}} \NOT(\pi{\uparrow}i),$ }  \label{alg2-counterexamples-init}
    found $\gets$ {\BatchSynt}($0$, $\textit{PS}^0,\dots,\textit{PS}^n$, $\phi$,$b$)\; \label{alg2-call-batchsynth}
    \lIf{$\textit{found} \neq \emptyset$}{\Return{found}}
}
\Return{$\emptyset$}\;
}
\Fn{\BatchSynt{$i,\textit{PS}^0,\dots,\textit{PS}^n, \phi, b$}}{
\KwResult{$r_0, \dots, r_n$ with $r_i \vDash \textit{PS}^i$ and
$r_0 \parallel \dots \parallel r_n \vDash \phi$ or $\emptyset$ otherwise}
$j \gets 0$\;
 \While{$\textit{PS}^i$ has unprocessed instances and $j < b$}{  \label{alg2-for-loop}
            $r_i \gets \text{next instance of } \textit{PS}^i$\;
            \If{$i=n$}{
                 \lIf{$r_0 \parallel \dots \parallel r_n \vDash \phi$}{\Return{$\{r_0,\dots,r_n \}$}}    \label{alg2-base-case}
                 $cexs \gets \text{ process counterexamples}$\;
              }
             \If{$i<n$}{
                $\textit{found} \gets {\BatchSynt}(i+1,\textit{PS}^0,\dots,\textit{PS}^n,\phi,b)$\; \label{alg2-recursive-call}
                \lIf{$\textit{found} \neq \emptyset$}{\Return{$\textit{found}$}}
            }
    $j \gets j+1$ \; 
}
\Return{$\emptyset$}\;
}

\end{algorithm*}
}

\section{Experimental Evaluation}\label{sec:examples}
{\tiny
\begin{table}[!ht]
\begin{tabular}{l l}
\begin{minipage}[b]{0.45\linewidth}
    \centering
    \vspace{-0.5cm}
    \begin{tabular}{|l|l|l|l|l|l|l|l|}
    \hline
        Ex. & Sc. & L.Time & G.Time & It. & R.St. & T.St. & Res. \\ \hline
        \textsf{mut(2)} & 4 & 0.222 & 0.461 & 9 & $2^{2.58}$ & $2^{9.33}$ & F \\ \hline
        \textsf{mut(3)} & 4 & 0.204 & 0.877 & 27 & $2^3$ & $2^{13.50}$ & F \\ \hline
        \textsf{mut(4)} & 4 & 0.206 & 2.945 & 81 & $2^{3.32}$ & $ 2^{17.67}$ & F \\ \hline
        \textsf{mut(5)} & 4 & 0.208 & 11.05 & 243 & $ 2^{3.58}$ & $ 2^{21.84}$ & F \\ \hline
        \textsf{mut(6)} & 4 & 0.209 & 47.634 & 729 & $ 2^{3.80}$ & $ 2^{26.01}$ & F \\ \hline
        \textsf{mut(7)} & 4 & 0.21 & 220.863 & 2187 & $ 2^4$ & $ 2^{30.18}$ & F \\ \hline
        \textsf{phil(2)} & 14 & 0.904 & 1.226 & 5  & $2^4$ & $2^{14.33}$ & F \\ \hline
        \textsf{phil(3)} & 14 & 0.82 & 1.857 & 14 & $ 2^{5.95}$ & $2^{21.50}$ & F \\ \hline
        \textsf{phil(4)} & 14 & 0.801 & 5.987 & 41 & $2^8$ & $2^{28.67}$ & F \\ \hline
        \textsf{phil(5)} & 14 & 0.81 & 197.652 & 335 & $2^{9.39}$ & $2^{35.84}$ & F \\ \hline
        \textsf{phil(6)} & 14 & 0.808 & 973.321 & 365 & $2^{12}$ & $2^{43.01}$ & F \\ \hline
        \textsf{phil(7)} & 14 & 0.867 & TO & - & - & - & TO \\ \hline
        \textsf{bar(2)} & 16 & 0.768 & 1.535 & 11 & $2^{2.58}$ & $2^{14.58}$ & F \\ \hline
        \textsf{bar(3)} & 16 & 1.056 & 10.761 & 97 & $2^{5.90}$ & $2^{23.16}$ & F \\ \hline
        \textsf{bar(4)} & 16 & 2.978 & 62.111 & 85 & $2^{10.75}$ & $2^{32.16}$ & F \\ \hline
        \textsf{bar(5)} & 16 & 3.876 & TO & - & - & - & TO \\ \hline
        \textsf{pet(2)} & 12 & 0.414 & 0.511 & 2 & $2^2$ & $2^{8.16}$ & F \\ \hline
        \textsf{pet(3)} & 19 & TO & - & - & - & - & TO \\ \hline
        \textsf{pet(3)} & 20 & 238.777 & 239.337 & 5 & $2^{3.32}$ & $2^{16.22}$ & F \\ \hline
        \textsf{r(1)w(1)} & 6 & 0.28 & 0.446 & 7 & $2^{1.58}$ & $2^7$ & F \\ \hline
       \textsf{r(1)w(2)} & 6 & 0.323 & 0.865 & 25 & $2^2$ & $2^{10}$ & F \\ \hline
        \textsf{r(1)w(3)} & 6 & 0.361 & 2.112 & 79 & $2^{2.32}$ & $2^{13}$ & F \\ \hline
        \textsf{r(1)w(4)} & 6 & 0.389 & 6.211 & 241 & $2^{2.58}$ & $2^{16}$ & F \\ \hline
        \textsf{r(1)w(5)} & 6 & 0.441 & 19.625 & 723 & $2^{2.80}$ & $2^{19}$ & F \\ \hline
        \textsf{r(1)w(6)} & 6 & 0.459 & 67.666 & 2181 & $2^3$ & $2^{22}$ & F \\ \hline
        \textsf{r(2)w(1)} & 12 & 0.644 & 1.531 & 29 & $2^{2.32}$ & $2^{13.16}$ & F \\ \hline
        \textsf{r(2)w(2)} & 12 & 0.898 & 6.981 & 167 & $2^{2.58}$ & $ 2^{16.16}$ & F \\ \hline
        \textsf{r(2)w(3)} & 12 & 1.173 & 109.623 & 1377 & $2^{2.80}$ & $2^{19.16}$ & F \\ \hline
        \textsf{r(2)w(4)} & 11 & - & 0.8 & - & - & - & U \\ \hline
        \textsf{r(2)w(4)} & 12 & 1.565 & TO & - & - & - & TO\\ \hline
        \textsf{r(3)w(1)} & 24 & 37.757 & 38.192 & 4 & $ 2^{3.16}$ & ${2^{13}}$ & F \\ \hline
        \textsf{r(3)w(2)} & 24 & 74.951 & 76.572 & 13 & $ 2^{3.32}$ & $2^{16}$ & F \\ \hline
        \textsf{r(3)w(3)} & 24 & 109.623 & 113.523 & 40 & $2^{3.45}$ & $2^{19}$ & F \\ \hline
        \textsf{r(3)w(4)} & 24 & 147.42 & 159.073 & 121 & $2^{3.58}$ & $2^{22}$ & F \\ \hline
        \textsf{r(3)w(5)} & 24 & 184.087 & 235.318 & 362 & $2^{3.70}$ & ${2^{25}}$ & F \\ \hline
        \textsf{r(3)w(6)} & 24 & 220.121 & 382.628 & 1091 & $2^{3.80}$ & ${2^{28}}$ & F \\ \hline
\end{tabular}
\vspace{0.1cm}
\caption{Results for \textsf{mut}, \textsf{phil}, \textsf{bar}, \textsf{pet}, and \textsf{rw} examples.}\label{tab:results-common}
\end{minipage}
&
\begin{minipage}{0.70\textwidth}
\vspace{-17cm}
\centering
\begin{tikzpicture}
	\begin{axis}[name=Mutex,height=4cm,width=5.5cm,
		xlabel=\scriptsize{\textsf{mut(n)}},
		ylabel=\scriptsize{Seconds},
		x label style={at={(axis description cs:0.5,0.1)},anchor=north},
    	    y label style={at={(axis description cs:0.07,1.1)},anchor=west,rotate=-90},
		ymax=900,
		clip=false,
		]
	\addplot[color=red,mark=*] coordinates {
		(2, 0.461)
		(3, 0.877)
		(4, 2.945)
		(5, 11.05)
		(6, 47.634)
		(7, 220.863)
	};
	\addplot[color=green,mark=*] coordinates {
		(2, 0.483)
		(3, 1.602)
		(4, 9.032)
		(5, 60.845)
		(6, 444.691)
	};
	\addplot[color=green,mark=*,dashed] coordinates {		
		(6, 444.691)
		(6.5, 960) 
	}node[pin={[pin distance=-0.1cm]90:{\tiny{TO}}}]{};
	
	\addplot[color=blue,mark=*] coordinates {
		(2,0.983)
		(3,4.693)
		(4,51.233)
		(5,683.642)
	};
	\addplot[color=blue,mark=*,dashed] coordinates {		
		(5,683.642)
		(5.3,960) 
	}node[pin={[pin distance=-0.1cm]90:{\tiny{TO}}}]{};
	
	\addplot[color=brown,mark=*] coordinates {
		(2,0.608)
		(3,5.016)
		(4,62.034)
	};
	\addplot[color=brown,mark=*,dashed] coordinates {
		(4,62.034)
		(4.5,960) 
	}node[pin={[pin distance=-0.1cm]90:{\tiny{TO}}}]{};
	\addplot[color=black,mark=*] coordinates {
		(2,0.348)
		(3,0.75)
		(4,2.679)
		(5,12.971)
		(6,65.403)
		(7,372.247)
	};
	\addplot[color=orange,mark=*] coordinates {
		(2,4.62)
		(3,4.664)
	};
	\addplot[color=orange,mark=*,dashed] coordinates {
		(3,4.664)
		(3.5,960)
	}node[pin={[pin distance=-0.1cm]90:{\tiny{TO}}}]{};
	
	\end{axis}
\end{tikzpicture}
\label{fig:comparison}
\begin{tikzpicture}
	\begin{axis}[name=Philosophers,height=4cm,width=5.5cm,
		xlabel=\scriptsize{\textsf{phil(n)}},
		ylabel=\scriptsize{Seconds},
		ymax=1100,
		clip=false,
		x label style={at={(axis description cs:0.5,0.1)},anchor=north},
    	y label style={at={(axis description cs:0.07,1.1)},anchor=west,rotate=-90},
		legend pos=north west]
	\addplot[color=red,mark=*] coordinates {
		(2,1.117)
		(3, 1.808)
		(4, 5.987)
		(5,197.652)
		(6,973.321)
	};
	\addplot[color=red,mark=*,dashed] coordinates {
		(6,973.321)
		(6.1,1160)
	}node[pin={[pin distance=-0.1cm]90:{\tiny{TO}}}]{};
	\addplot[color=green,mark=*] coordinates {
		(2,1.253)
		(3, 2.765)
		(4, 17.661)
	};
	\addplot[color=green,mark=*,dashed] coordinates {
		(4, 17.661)
		(4.5,1150)
	}node[pin={[pin distance=-0.1cm]60:{\tiny{TO}}}]{};
	
	\addplot[color=blue,mark=*] coordinates {
		(2,1.37)
		(3,5.552)
		(4,95.327)
	};
	\addplot[color=blue,mark=*,dashed] coordinates {
		(4,95.327)
		(4.5,1150)
	}node[pin={[pin distance=-0.1cm]90:{\tiny{TO}}}]{};
	\addplot[color=brown,mark=*] coordinates {
		(2,1.37)
		(3,7.264)
		(4,142.584)
	};
	\addplot[color=brown,mark=*,dashed] coordinates {
		(4,142.584)
		(4.5,1150)
	}node[pin={[pin distance=-0.1cm]120:{\tiny{TO}}}]{};
	\addplot[color=black,mark=*] coordinates {
		(2,35.998)
		(3,97.636)
		(4,324)
	};
	\addplot[color=black,mark=*,dashed] coordinates {
		(4,324)
		(4.3,1150)
	}node[pin={[pin distance=-0.1cm]180:{\tiny{TO}}}]{};
	\addplot[color=orange,mark=*] coordinates {
		(2,5.916)
		(3,6.273)
		(4,12.845)
		 (5,  68.5)
	};
	\addplot[color=orange,mark=*, dashed] coordinates {
		 (5,  68.5)
	     (5.5,1150)
	}node[pin={[pin distance=-0.1cm]90:{\tiny{TO}}}]{};
	\end{axis}
\end{tikzpicture}
\begin{tikzpicture}
	\begin{axis}[name=SenseBarrier,height=4cm,width=5.5cm,
		xlabel=\scriptsize{\textsf{bar(n)}},
		ylabel=\scriptsize{Seconds},
		ymax=1200,
		clip=false,
		x label style={at={(axis description cs:0.5,0.1)},anchor=north},
    	y label style={at={(axis description cs:0.07,1.1)},anchor=west,rotate=-90},
		legend pos=north west]
	\addplot[color=red,mark=*] coordinates {
	    
		(2,1.535)
		(3, 10.761)
		(4, 62.111)
		(5,100)
		
	};
	\addplot[color=green,mark=*] coordinates {
	
		(2,1.672)
		(3, 10.569)
		(4, 436.365)
	};
	\addplot[color=green,mark=*,dashed] coordinates {
		(4, 436.365)
		(4.5,1260) 
	}node[pin={[pin distance=-0.1cm]90:{\tiny{TO}}}]{};
	
	\addplot[color=blue,mark=*] coordinates {
		(2,2.304)
	};
	\addplot[color=blue,mark=*,dashed] coordinates {
		(2,2.304)
		(2.5,1260)
	}node[pin={[pin distance=-0.1cm]90:{\tiny{TO}}}]{};
	\addplot[color=brown,mark=*] coordinates {
	
		(2,2.362)
		(3,37.755)
	};
	\addplot[color=brown,mark=*,dashed] coordinates {
	
		(3,37.755)
		(3.5,1260) 
	}node[pin={[pin distance=-0.1cm]90:{\tiny{TO}}}]{};
	\addplot[color=black,mark=*] coordinates {
	   
		(2,86.916)
	};
	\addplot[color=black,mark=*,dashed] coordinates {
	   
		(2,86.916)
		(2.5,1260)
	}node[pin={[pin distance=-0.1cm]90:{\tiny{TO}}}]{};
	\addplot[color=orange,mark=*] coordinates {
		(2,4.233)
		(3, 5.31)
		(4,1140.9)
	};
	\addplot[color=orange,mark=*,dashed] coordinates {
		(4,1140.9)
		(4.05,1260)
	}node[pin={[pin distance=-0.1cm]90:{\tiny{TO}}}]{};
	\end{axis}
\end{tikzpicture}

\begin{tikzpicture}
	\begin{axis}[name=Readers and Writers,height=4cm,width=5.5cm,
		xlabel=\scriptsize{\textsf{r(1)w(n)}},
		ylabel=\scriptsize{Seconds},
		x label style={at={(axis description cs:0.5,0.1)},anchor=north},
    	y label style={at={(axis description cs:0.07,1.1)},anchor=west,rotate=-90},
    	ymax=600,
		clip=false,
		legend style={at={(0.5,-0.3)},anchor=north,legend  columns =3, transpose legend}]
     \addlegendimage{red, line legend, mark=*} 
    \addlegendentry{\textsf{exp2}}
    \addlegendimage{green, line legend, , mark=*} 
    \addlegendentry{\textsf{exp4}}
    \addlegendimage{blue, line legend, , mark=*} 
    \addlegendentry{\textsf{exp8}}
    \addlegendimage{brown, line legend, , mark=*} 
    \addlegendentry{\textsf{lineal10}}
    \addlegendimage{black,,line legend,  mark=*} 
    \addlegendentry{\textsf{nocex}}
    \addlegendimage{orange,,line legend,  mark=*} 
    \addlegendentry{\textsf{PSketch}}
	\addplot[color=red,mark=*] coordinates {
		(1,0.492)
		(2,0.813)
		(3,1.603)
		(4,4.069)
		(5,11.705)
		(6,37.703)
	};
	\addplot[color=green,mark=*] coordinates {
		(2,0.442)
		(3,4.055)
		(4,19.064)
		(5,103.635)
		(6,590.277)
	};
	\addplot[color=blue,mark=*] coordinates {
		(1,0.625)
		(2,2.247)
		(3,16.51)
		(4,161.499)
	};
	\addplot[color=blue,mark=*,dashed] coordinates {
		(4,161.499)
		(4.7,650) 
	}node[pin={[pin distance=-0.1cm]130:{\tiny{TO}}}]{};
	\addplot[color=brown,mark=*] coordinates {
		(1,0.654)
		(2,2.892)
		(3,28.208)
		(4,348.931)
	};
	\addplot[color=brown,mark=*,dashed] coordinates {
		(4,348.931)
		(4.3,650) 
	}node[pin={[pin distance=-0.1cm]180:{\tiny{TO}}}]{};
	\addplot[color=black,mark=*] coordinates {
		(2,1.344)
		(3,14.807)
		(4,93.844)
	};
	\addplot[color=black,mark=*,dashed] coordinates {
		(4,93.844)
		(5,650) 
	}node[pin={[pin distance=-0.1cm]90:{\tiny{TO}}}]{};
	
	\addplot[color=orange,mark=*] coordinates {
		(1,8.465)
		(2,8.483)
		(3,10.463)
		(4,33.898)
		(5,42.824)
	};
	\addplot[color=orange,mark=*,dashed] coordinates {
		(5,42.824)
		(5.5,650)
	}node[pin={[pin distance=-0.1cm]70:{\tiny{TO}}}]{};
	\end{axis}
\end{tikzpicture}
\vspace{0.4cm}
\captionof{figure}{Comparison between \\ \textsf{exp2}, \textsf{exp4}, \textsf{exp8},  \textsf{lineal10},  \textsf{nocex}, \\ and {\PSketch}.}
\label{fig:examples-plot}
\end{minipage}
\end{tabular}
\vspace{-0.5cm}
\end{table}
}
We evaluate our approach around the following research questions: 
\begin{description}
\item[RQ1] \emph{How effective/efficient is our synthesis approach?}
\item[RQ2] \emph{How good is the counterexample-guided search for speeding up the synthesis method?}
\item[RQ3] \emph{How do the selected bounds affect the synthesis method?}
\item[RQ4] \emph{How does our approach compare with related approaches?}
\end{description}
To answer these questions,  we implemented Alg.~\ref{alg:improved_alg} in a prototype tool, it uses the \textsf{Alloy Analyzer}~\cite{AlloyBook} for obtaining instances of specifications, and {\NuSMV}~\cite{Cimatti+2002} for model checking the candidates.  We evaluate our approach on eight  examples of distributed algorithms: the dining philosophers (\textsf{phil})~\cite{Dijkstra71} (our running example), Mutex (\textsf{mut})~\cite{Fokkink13},  Readers and Writers (\textsf{rw})~\cite{Fokkink13}, the generalized version of Peterson's algorithm (\textsf{pet})~\cite{Fokkink13},   and the combined-tree Barrier protocol (\textsf{bar})~\cite{Fokkink13}.  Furthermore,  we  also encoded the arbiter examples presented in \cite{Party,Piterman+2006}: a simple arbiter (\textsf{arb}), a full arbiter (\textsf{farb}), and the Pnueli arbiter (\textsf{parb}). These case studies assume a distributed token ring architecture,  which we modeled using the Alloy language.  

Tables~\ref{tab:results-common} and~\ref{tab:results-arbiter}  summarize the experimental results. The experiments were conducted on an Apple M2 processor with 16GB of memory.   In these examples, we used the sequence of bounds $2,4,8,16,\dots$, named \textsf{exp2} from now on.
For each case study, we report the 
bound over the size of the processes (Sc), the maximum time needed to generate local process instances (L.Time), the total time required for synthesizing the system (T.Time), the number of times that the model checker was invoked (Its) by the synthesizer, and the final result of our synthesis algorithm:  `F' (if an implementation was found),  `N'  (if no implementation was found),  `TO'  if the example timed out, or `U'  (if the specification was found unsatisfiable by the Alloy tool).  We also report the number of reachable states (R.St.), and the total states (T.St.) of the obtained implementations, expressed as power of $2$.  For space reasons, we only include a few configurations found unsatisfiable by the solver; similar numbers can be obtained for the rest of the cases if the specifications are processed with smaller scopes. We also indicate the number of processes considered in each experiment.  For instance,  \textsf{phil(6)} indicates that we considered \textsf{6} concurrent processes (i.e., philosophers) in the dining philosophers example. In the case of Readers and Writers, 
\textsf{r(n)w(m)} means that \textsf{n} readers and \textsf{m} writers were considered.  We  have  set out a time out of 30 minutes.
\begin{wraptable}[19]{r}{.60\textwidth}
\vspace{-0.5cm}
\begin{tabular}{|l|l|l|l|l|l|l|l|}
    \hline
        Ex. & Sc. & L.Time & G.Time & It. & R.St. & T.St. & Res. \\ \hline
        \textsf{arb(2)} & 12 & 0.814 &  1.322 & 7 & $2^{2.32}$  & $2^{12}$ & F \\ \hline
        \textsf{arb(3)} & 12 & 0.728 & 2.11 & 16 & $2^4$ & $2^{18}$ & F \\ \hline
        \textsf{arb(4)} & 12 & 0.854 & 4.303 & 16 & $2^{5.24}$ & $2^{24}$ & F \\ \hline
        \textsf{arb(5)} & 12 & 0.976 & 12.54 & 16 & $2^{6.35}$ & $2^{30}$ & F \\ \hline
        \textsf{arb(6)} & 12 & 1.134 & 52.344 & 16 & $2^{7.40}$ & $2^{36}$ & F \\ \hline
        \textsf{farb(2)} & 12 & 0.552 & 1.164 & 8 & $2^{2.58}$ & $2^{12}$ & F \\ \hline
        \textsf{farb(3)} & 12 & 0.75 & 2.97 & 20 & $2^{3.45}$ & $2^{18}$ & F \\ \hline
        \textsf{farb(4)} & 12 & 0.935 & 15.602 & 56 & $2^{4.39}$ & $2^{24}$ & F \\ \hline
        \textsf{farb(5)} & 12 & 1.232 & 85.933 & 164 & $2^{5.35}$ & $2^{30}$ & F \\ \hline
        \textsf{farb(6)} & 12 & 2.034 & 672.539 & 488 & $2^{6.33}$ & $2^{36}$ & F\\ \hline
        \textsf{parb(2)} & 12 & 0.662 & 1.166 & 8 & $2^{2.80}$ & $2^{12}$ & F\\ \hline
        \textsf{parb(3)} & 12 & 0.714 & 2.277 & 20 & $2^{3.45}$ & $2^{18}$ & F \\ \hline
        \textsf{parb(4)} & 12 & 0.886 & 8.593 & 56 & $2^{4.39}$ & $2^{24}$ & F\\ \hline
        \textsf{parb(5)} & 12 & 1.092 & 54.25 & 164 & $2^{5.35}$ & $2^{30}$ & F \\ \hline
        \textsf{parb(6)} & 12 &1.324 & 700.731 & 488 & $2^{6.33}$ & $2^{36}$ & F \\ \hline
\end{tabular}
\caption{Results for the arbiter examples.}\label{tab:results-arbiter}
\end{wraptable}
Table \ref{tab:results-common} shows that the technique scales reasonably well for those case studies in which each process uses only a reduced number of locks and shared variables (e.g., dining philosophers, mutex and reader-writers). For the cases where the number of shared variables accessed by the processes is bigger (e.g., Peterson for $n$ processes), the technique does not scale that well. Intuitively, more shared variables imply more actions performed by the environment, which increases the size of the formula fed to the SAT solver.  We plan to investigate how to equip specifications with assumptions on the environment's behavior, to restrict the possible values of shared variables; this may simplify the SAT problem when searching for local implementations.  It is worth noting, that even though the algorithm is incomplete, we have not 
observed any ``not found'' outputs in our benchmarks. This could be due to  the set timeouts.  We leave an in-depth investigation of this as further work.

\begin{wrapfigure}[31]{hr}{.40\textwidth}
\vspace{-1.5cm}
\begin{tikzpicture}
	\begin{axis}[name=Arbiter,height=4cm,width=5.5cm,
		xlabel=\scriptsize{Arbiter(n)},
		ylabel=\scriptsize{Seconds},
		x label style={at={(axis description cs:0.5,0.1)},anchor=north},
    	y label style={at={(axis description cs:0.07,1.1)},anchor=west,rotate=-90},
    	ymax=400,
		clip=false,
		legend pos=north west]
	\addplot[color=red,mark=*] coordinates {
		(2,1.3)
		(3,2.11)
		(4,4.3)
		(5,12.54)
		(6,52.34)
	};
	\addplot[color=green,mark=*] coordinates {
		(2,1.372)
		(3,3.545)
		(4,8.824)
		(5,26.531)
		(6,107.897)
	};
	\addplot[color=blue,mark=*] coordinates {
		(2,1.72)
		(3,7.022)
		(4,18.176)
		(5,63.028)
		(6, 288.419) 
	};
	\addplot[color=brown,mark=*] coordinates {
		(2,1.737)
		(3,9.11)
		(4,24.866)
		(5,88.064)
		(6,382.565)
	};
	\addplot[color=pink,mark=*] coordinates {
		(2,0.63)
		(3,1.18)
		(4,2.82)
		(5,6.12)
		(6,12.43)
	};
	\addplot[color=black,mark=*] coordinates {
		(2,440)
	}node[pin={[pin distance=-0.1cm]90:{\tiny{TO}}}]{};

	\end{axis}
\end{tikzpicture}
\begin{tikzpicture}
	\begin{axis}[name=FullArbiter,height=4cm,width=5.5cm,
		xlabel=\scriptsize{FullArbiter(n)},
		ylabel=\scriptsize{Seconds},
		x label style={at={(axis description cs:0.5,0.1)},anchor=north},
    	y label style={at={(axis description cs:0.07,1.1)},anchor=west,rotate=-90},
    	ymax=1200,
		clip=false,
		legend pos=north west]
	\addplot[color=red,mark=*] coordinates {
		(2,1.083)
		(3,2.774)
		(4,12.873)
		(5,82.981)
		(6,700.731)
	};
	\addplot[color=green,mark=*] coordinates {
		(2,1.24)
		(3,5.719)
		(4,34.7)
		(5,309.953)
	};
	\addplot[color=green,mark=*,dashed] coordinates {
		(5,309.953)
		(6,1250)
	}node[pin={[pin distance=-0.1cm]90:{\tiny{TO}}}]{};
	
	\addplot[color=blue,mark=*] coordinates {
		(2,1.55)
		(3,12.382)
		(4,145.86)
	};
	\addplot[color=blue,mark=*,dashed] coordinates {
		(4,145.86)
		(5,1250)
	}node[pin={[pin distance=-0.1cm]90:{\tiny{TO}}}]{};
	
	\addplot[color=brown,mark=*] coordinates {
		(2,1.677)
		(3,15.45)
		(4,232.942)
	};
	\addplot[color=brown,mark=*,dashed] coordinates {
		(4,232.942)
		(4.8,1250)
	}node[pin={[pin distance=-0.1cm]120:{\tiny{TO}}}]{};
	    
	\addplot[color=pink,mark=*] coordinates {
		(2,0.68)
		(3,1.43)
		(4,4.078)
		(5,8.85)
		(6,32.73)
	};
	\addplot[color=black,mark=*,dashed] coordinates {
		(2,196.115)
		(3,1260) 
	}node[pin={[pin distance=-0.1cm]90:{\tiny{TO}}}]{};
	
	\end{axis}
\end{tikzpicture}
\begin{tikzpicture}
	\begin{axis}[name=PnueliArbiter,height=4cm,width=5.5cm,
		xlabel=\scriptsize{PnueliArbiter(n)},
		ylabel=\scriptsize{Seconds},
		x label style={at={(axis description cs:0.5,0.1)},anchor=north},
    	y label style={at={(axis description cs:0.07,1.1)},anchor=west,rotate=-90},
    	ymax=1400,
		clip=false,
		legend style={at={(0.5,-0.3)},anchor=north,legend  columns =3, transpose legend}]
     \addlegendimage{red, line legend, mark=*} 
    \addlegendentry{\textsf{exp2}}
    \addlegendimage{green, line legend, , mark=*} 
    \addlegendentry{\textsf{exp4}}
    \addlegendimage{blue, line legend, , mark=*} 
    \addlegendentry{\textsf{exp8}}
    \addlegendimage{brown, line legend, , mark=*} 
    \addlegendentry{\textsf{lineal10}}
    \addlegendimage{black,,line legend,  mark=*} 
    \addlegendentry{\textsf{nocex}}
    \addlegendimage{pink,line legend,  mark=*} 
    \addlegendentry{\textsf{Party}}
	\addplot[color=red,mark=*] coordinates {
		(2,1.206)
		(3,2.674)
		(4,8.639)
		(5,59.893)
		(6,905.203) 
	};
	\addplot[color=green,mark=*] coordinates {
		(2,1.406)
		(3,5.461)
		(4,29.063)
		(5,274.013)
	};
	\addplot[color=green,mark=*,dashed] coordinates {
		(5,274.013)
		(5.4,1450) 
	}node[pin={[pin distance=-0.1cm]110:{\tiny{TO}}}]{};
	
	\addplot[color=blue,mark=*] coordinates {
		(2,1.634)
		(3,9.749)
		(4,128.565)
	};
	\addplot[color=blue,mark=*,dashed] coordinates {
		(4,128.565)
		(4.6, 1450) 
	}node[pin={[pin distance=-0.1cm]90:{\tiny{TO}}}]{};
	\addplot[color=brown,mark=*] coordinates {
		(2,1.789)
		(3,13.855)
		(4,201.653)
	};
	\addplot[color=brown,mark=*,dashed] coordinates {	
		(4,201.653)
		(4.5,1450)
	}node[pin={[pin distance=-0.1cm]130:{\tiny{TO}}}]{};
	\addplot[color=black,mark=*,dashed] coordinates {
		(2,246.157)
	 	(2.5,1450)
	}node[pin={[pin distance=-0.1cm]90:{\tiny{TO}}}]{};
	\addplot[color=pink,mark=*] coordinates {
		(2,0.69)
		(3,1.162)
		(4,2.408)
		(5,11.836)
	};
	\addplot[color=pink,mark=*,dashed] coordinates {
		(5,11.836)
		(5.5,1450)
	}node[pin={[pin distance=-0.1cm]80:{\tiny{TO}}}]{};
	\end{axis}
\end{tikzpicture}
\caption{Efficiency comparison for arbiter examples}\label{fig:arbiter-plots}
\end{wrapfigure}
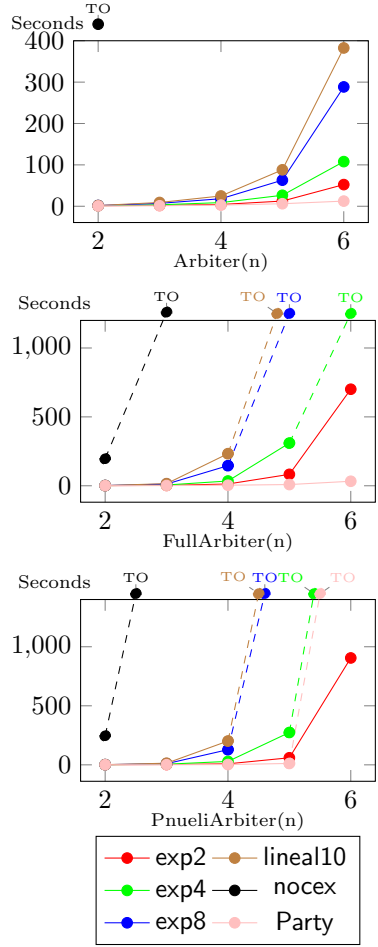

 To answer \textbf{RQ3}   we compare the results obtained with several configurations of bounds for the exploration phase, namely: \textsf{exp4} ($4,16,64,\dots$),  \textsf{exp8} ($8,64,512,\dots$),  \textsf{lineal10} ($10,20,30,\dots$), and for \textbf{RQ2} we also considered Alg.~\ref{alg:the_alg},  which does not take into account counterexamples (\textsf{nocex}).  The obtained results are depicted in Figs.\ref{fig:examples-plot} and \ref{fig:arbiter-plots}.  Note that time outs are remarked using dashed lines going out of the $y$-axis.
 In general, \textsf{exp2} behaves better than the other options,  thus it seems better to collect a few counterexamples first, and use them to improve the search.  A possible drawback of this setting is that a wrong choice of the first counterexamples may have as a consequence that no implementation is found,  i.e., one may expect that this configuration  is ``more incomplete''   than the other options. However, we have not observed this in our benchmarks.  Note that \textsf{nocex} timed out in many examples. Indeed, for the arbiter examples, \textsf{nocex} was able only to solve the examples with two processes,  taking for that more than $7000$ iterations.

To answer \textbf{RQ4} we have included in out analysis the synthesis tool {\PSketch}~\cite{Solar-Lezama+2008}, that implements a Counterexample-driven Guided Inductive Synthesis (CEGIS) algorithm to obtain code from sketched code (i.e., code annotated with ``holes''). To run {\PSketch}, we took the Dining Philosophers specification provided in~\cite{Solar-Lezama+2008}, and manually elaborated the specification for Mutex,  Readers-Writers, and the Barrier example.  The Peterson example cannot be analysed with  {\PSketch} since this tool only supports the analysis of safety properties.  For the arbiter case studies, we have compare against the tool {\Party} \cite{Party} which is a tool specifically tailored for distributed systems that use token ring architectures.

Our comparison focuses only on the time required by each technique for synthesizing the distributed solutions. The plots of Figs.~\ref{fig:examples-plot} and \ref{fig:arbiter-plots} depict the results of this comparison. In all the case studies, we notice that the efficiency of {\PSketch} is drastically affected as the number of processes to synthesize is incremented. For instance, in the dining philosophers with 6 processes, {\Sketch} timed out; in contrast, our tool was able to obtain a solution. Similarly, in the case of 1 reader and 6 writers, {\PSketch} failed in synthesizing a solution, while our approach succeeded. A similar analysis applies to Mutex.

  In the case of the tool {\Party}, for the \textsf{arb} and \textsf{farb} examples {\Party} was able to find solutions faster than \textsf{exp2}; it must be noted that in these cases {\Party} uses a cut-off of $4$,  i.e., it reduces configurations with $n>4$ processes to the case $n=4$.  However,  even though {\Party} has several optimizations for token ring systems, our tool was able to synthesize an implementation of the \textsf{parb(6)} and {\Party} timed out for this case. 
  
It is worth noting that the tool can be used to find different solutions for some examples.  We have experimented with this using the \textsf{phil} example, where, after finding an initial solution, we allow the algorithm to keep looking for further solutions, and it succeeded in finding a second solution.  For space reasons we do not investigate this aspect of the tool further here.


\section{Related Work}\label{sec:related}	

The seminal work of Emerson and Clarke \cite{EmersonClarke82}, and related approaches \cite{Attie16,EmersonSamanta11}, are based on tableau methods for synthesizing synchronization skeletons from global {\CTL} specifications,  which are impractical in most cases, or require to provide process descriptions as state machines.
The synthesis of distributed reactive systems \cite{PnueliRosner90}  aim to synthesize a distributed architecture that reacts to an environment when possible. This is a very general problem and can be undecidable in many cases.  Bounded synthesis \cite{DBLP:journals/sttt/FinkbeinerS13} focuses on reactive (synchronous) systems,
or assumes a \emph{causal memory}  model \cite{DBLP:journals/iandc/FinkbeinerO17},  i.e.,  the components memorize their histories and synchronize according to this.  Also, we remark that all these approaches consider only global {\LTL} specifications,  in our approach we also consider local specifications in a pre/post-condition and invariant style, which may include Alloy formulas.

Counterexample-driven guided inductive synthesis (CEGIS) \cite{Abate+2018} reduces the problem of finding a given program to solve a (bounded) first-order formula; then inductive generalization and verifiers are used for generating candidate solutions. {\AlloyStar}  \cite{Milicevic+2015} implements a general version of CEGIS. In contrast to CEGIS, our approach starts from ``loose'' versions of the processes, which are refined until a solution is obtained. Thus, we can accumulate counterexamples, and use them throughout the synthesis process.  As noted in \cite{Solar-Lezama+2008},  
using executions of asynchronous concurrent programs as counterexamples in  CEGIS is not direct.  Particularly, {\Sketch} is only able to handle safety properties.  Other works, e.g. \cite{VechevYY13,Party}, consider only safety properties, or are designed for specific settings.

\section{Final Remarks}
\label{sec:conclusions}

We presented a correct-by-construction bounded approach for the synthesis of distributed algorithms. The approach iteratively generates candidate process implementations using the Alloy tool and manipulates the found counterexamples to extract information that helps to refine  the search. In contrast to other approaches, our synthesis algorithm performs a local reasoning, to avoid the explicit construction of the global state space, and applies to general temporal properties, including safety and liveness. We developed a prototype tool that effectively solved common case studies of distributed algorithms. As noted in Section~\ref{sec:examples} the number of shared variables may affect the scalability of our algorithm. In view of this, we plan to extend our approach in several ways. e.g.,  exploring other heuristics to improve the exploration of the instance space.  One benefit of the proposed approach is the versatility provided by the Alloy notation, which allows us to model different kinds of distributed systems, as illustrated with the token ring systems in Section~\ref{sec:examples}.



\bibliographystyle{splncs}
\bibliography{myBib}

\begin{thebibliography}{10}

\bibitem{MannaWolper84}
Manna, Z., Wolper, P.:
\newblock Synthesis of communicating processes from temporal logic.
\newblock ACM Trans. Program. Lang. Syst \textbf{6} (1984)

\bibitem{PnueliRosner89}
Pnueli, A., Rosner, R.:
\newblock On the synthesis of a reactive module.
\newblock In ACM, ed.: Principles of Programming Languages. (1989)

\bibitem{EmersonClarke82}
Emerson, E.A., Clarke, E.M.:
\newblock Using branching time temporal logic to synthesize synchronization
  skeletons.
\newblock Sci. Comput. Program. \textbf{2}(3) (1982)

\bibitem{BGJPPW07a}
Bloem, R., Galler, S., Jobstmann, B., Piterman, N., Pnueli, A., Weiglhofer, M.:
\newblock Automatic hardware synthesis from specifications: A case study.
\newblock In: Design Automation and Test in Europe, acm (2007)

\bibitem{Piterman+2006}
Piterman, N., Pnueli, A., Sa'ar, Y.:
\newblock Synthesis of reactive(1) designs.
\newblock In Emerson, E.A., Namjoshi, K.S., eds.: Verification, Model Checking,
  and Abstract Interpretation, 7th International Conference, {VMCAI} 2006,
  Charleston, SC, USA, January 8-10, 2006, Proceedings. Volume 3855 of Lecture
  Notes in Computer Science., Springer (2006)  364--380

\bibitem{Jha+2010}
Jha, S., Gulwani, S., Seshia, S.A., Tiwari, A.:
\newblock Oracle-guided component-based program synthesis.
\newblock In: ICSE 2010, New York, NY, USA, Association for Computing Machinery
  (2010)

\bibitem{Abate+2018}
Abate, A., David, C., Kesseli, P., Kroening, D., Polgreen, E.:
\newblock Counterexample guided inductive synthesis modulo theories.
\newblock In Chockler, H., Weissenbacher, G., eds.: Computer Aided
  Verification, Cham, Springer International Publishing (2018)

\bibitem{Srivastava+2010}
Srivastava, S., Gulwani, S., Foster, J.S.:
\newblock From program verification to program synthesis.
\newblock SIGPLAN Not. \textbf{45}(1) (January 2010)

\bibitem{PnueliRosner90}
Pnueli, A., Rosner, R.:
\newblock Distributed reactive systems are hard to synthesize.
\newblock In: Proceedings of the 31st Annual Symposium on Foundations of
  Computer Science. SFCS '90, USA, IEEE Computer Society (1990)

\bibitem{AlloyBook}
Jackson, D.:
\newblock Software Abstractions: Logic, Language and Analysis.
\newblock MIT Press (2016)

\bibitem{Katoen08}
Baier, C., Katoen, J.P.:
\newblock Principles of Model Checking.
\newblock MIT Press (2008)

\bibitem{PeledWilke1997}
Peled, D., Wilke, T.:
\newblock Stutter-invariant temporal properties are expressible without the
  next-time operator.
\newblock Information Processing Letters \textbf{63}(5) (1997)

\bibitem{Dijkstra1975}
Dijkstra, E.W.:
\newblock Guarded commands, non-determinacy and formal derivation of programs.
\newblock Comm. ACM \textbf{18}(8) (1975)

\bibitem{AroraGouda93}
Arora, A., Gouda, M.:
\newblock Closure and convergence: A foundation of fault-tolerant computing.
\newblock IEEE Transactions on Software Engineering \textbf{19}(11) (1993)

\bibitem{Dijkstra71}
Dijkstra, E.W.:
\newblock Hierarchical ordering of sequential processes.
\newblock Acta Informatica \textbf{1}(2) (1971)

\bibitem{Biere+1999}
Biere, A., Cimatti, A., Clarke, E.M., Zhu, Y.:
\newblock Symbolic model checking without bdds.
\newblock In: Proc. of 5th International Conference on Tools and Algorithms for
  Construction and Analysis of Systems. TACAS '99, Springer-Verlag (1999)

\bibitem{Cimatti+2002}
Cimatti, A., Clarke, E., Giunchiglia, E., Giunchiglia, F., Pistore, M., Roveri,
  M., Sebastiani, R., Tacchella, A.:
\newblock {NuSMV Version 2: An OpenSource Tool for Symbolic Model Checking}.
\newblock In: CAV 2002, Copenhagen, Denmark, Springer (July 2002)

\bibitem{Fokkink13}
Fokkink, W.:
\newblock Distributed Algorithms: An Intuitive Approach.
\newblock The MIT Press (2013)

\bibitem{Party}
Khalimov, A., Jacobs, S., Bloem, R.:
\newblock {PARTY} parameterized synthesis of token rings.
\newblock In Sharygina, N., Veith, H., eds.: Computer Aided Verification - 25th
  International Conference, {CAV} 2013, Saint Petersburg, Russia, July 13-19,
  2013. Proceedings. Volume 8044 of Lecture Notes in Computer Science.,
  Springer (2013)  928--933

\bibitem{Solar-Lezama+2008}
Solar{-}Lezama, A., Jones, C.G., Bod{\'{\i}}k, R.:
\newblock Sketching concurrent data structures.
\newblock In: Proceedings of the {ACM} {SIGPLAN} 2008 Conference on Programming
  Language Design and Implementation, Tucson, AZ, USA, June 7-13, 2008. (2008)

\bibitem{Attie16}
Attie, P.C.:
\newblock Synthesis of large dynamic concurrent programs from dynamic
  specifications.
\newblock Formal Methods in System Design \textbf{48}(1-2) (2016)

\bibitem{EmersonSamanta11}
Emerson, E.A., Samanta, R.:
\newblock An algorithmic framework for synthesis of concurrent programs.
\newblock In: Automated Technology for Verification and Analysis, 9th
  International Symposium, {ATVA} 2011, Taipei, Taiwan, October 11-14, 2011.
  Proceedings. Volume 6996 of LNCS., Springer (2011)

\bibitem{DBLP:journals/sttt/FinkbeinerS13}
Finkbeiner, B., Schewe, S.:
\newblock Bounded synthesis.
\newblock Int. J. Softw. Tools Technol. Transf. \textbf{15}(5-6) (2013)
  519--539

\bibitem{DBLP:journals/iandc/FinkbeinerO17}
Finkbeiner, B., Olderog, E.:
\newblock Petri games: Synthesis of distributed systems with causal memory.
\newblock Inf. Comput. \textbf{253} (2017)  181--203

\bibitem{Milicevic+2015}
Milicevic, A., Near, J.P., Kang, E., Jackson, D.:
\newblock Alloy*: {A} general-purpose higher-order relational constraint
  solver.
\newblock In: Proc. of 37th {IEEE/ACM} International Conference on Software
  Engineering ICSE 2015. (2015)

\bibitem{VechevYY13}
Vechev, M.T., Yahav, E., Yorsh, G.:
\newblock Abstraction-guided synthesis of synchronization.
\newblock {STTT} \textbf{15}(5-6) (2013)

\end{thebibliography}


\appendix
\section*{Appendix}
\noindent \textbf{Proof of Theorem  \ref{theorem:stuttering-equiv}. (Sketch)} We prove that for any trace $\pi' \in \traces{T^0 \parallel \dots \parallel T^k}$ starting at state $s$
we have that the projection of this trace to process $T^i$ is stutter equivalent to some trace $\pi \in \traces{T^i}$ starting at state $(s \proj i)$. 
Since stutter equivalence preserves $\LTLX$ properties \cite{PeledWilke1997},  the result follows. 

	Let $\pi' \in \traces{T^0 \parallel \dots \parallel T^k}$ starting a state $s$, then we prove that the trace $\pi \proj i = \pi[0] \proj i \rightarrow \pi[1] \proj i \rightarrow \dots$ is stutter equivalent to a trace 
$\pi \in \traces{T^i}$ starting at state $(s \proj i )$. $\pi$ is just defined by removing all the transitions in $\pi[j] \proj i \rightarrow \pi[j+1] \proj i$  that
do not correspond to transitions in $T^i$, it is simple to see that the removed transitions are stuttering steps, that is $L^i(\pi[j] \proj i)  = L^i(\pi[j+1] \proj i)$, otherwise if $L^i(\pi[j] \proj i)  \neq L^i(\pi[j+1] \proj i)$
the transition must correspond to the transition of another process (say $T^j$); so, $L^i(\pi[j] \proj i)$ and $L^i(\pi[j+1] \proj i)$ can only differ on their valuation of shared variables (they must coincide on the valuation of $\textit{Loc}^i$), but by the conditions \ref{system-spec-formula1}-\ref{system-spec-formula6} we have a matching transition in $T^i$ which is a contradiction. Also note that removing these transitions keeps the trace infinite, since $\pi$ is fair.
	Since we have removed only stuttering steps, the resulting execution is stutter equivalent to $\pi \proj i$. Now, given any {\LTLX} property $\phi$ then suppose
that $T^i \vDash \phi$ and $T^0 \parallel \dots \parallel T^k \nvDash  \phi$, that is, we have some $\pi \in \traces{T^0 \parallel \dots \parallel T^k}$ starting at $s$ such that
$\pi \nvDash \phi$ but since all the propositional variables in $\phi$ are in $Sh \cup Loc^i$ we have that $\pi \proj i \nvDash \phi$, and by the property described above,
we have a $\pi' \in \traces{T^i}(s \proj i)$ such that $\pi' \nvDash \phi$ which is a contradiction, and so $T^0 \parallel \dots \parallel T^k \vDash  \phi$.

\noindent  \textbf{Proof of Theorem \ref{theorem:preserve-safety}.  (Sketch)} 
	Observe that for any $\pi=s_0,s_1,\dots \in \traces{T'}$ such that $T' \vDash \refin{T,S}$ we have some $\pi' \in \traces{T}$ with $L(\pi) = L(\pi')$. Now assume that there is
some safety property such that $T' \nvDash \phi$ for some $T' \vDash \refin{T}{S}$ thus since $\phi$ is  safety property, we have a finite path $L(s_0),\dots,L(s_k)$ in which $\phi$
is not true, but then there is some $\pi'$ with the same prefix thus $T \nvDash \phi$ which is a contradiction.	
	
\noindent \textbf{Proof of Theorem 3. (Sketch)} It follows by the definition of $\NOT{(c)}$, note that $\NOT{(c)}  \equiv \neg \CNF{(c)}$, then 
if $T \vDash \refin{T, S}\oplus \NOT{(c)}$ then $T \nvDash \CNF{(c)}$, and the proof follows.

\noindent \textbf{Proof of Theorem 4 (Sketch)} Similar to Theorem 3.

\end{document}